\documentclass[12pt]{article}
\usepackage{url}
\usepackage{amsfonts}
\usepackage{amsmath,amssymb,color}
\usepackage{graphicx}
\usepackage{enumitem}
\usepackage{subcaption}
\usepackage{multirow}
\usepackage{bm}
\usepackage{apacite}
\usepackage{verbatim}
\usepackage{float}
\usepackage[toc,page]{appendix}
\usepackage[onehalfspacing]{setspace}
\usepackage{setspace}
\usepackage{natbib}

\usepackage[normalem]{ulem}

\newcommand{\yc}[1]{{{\color{black}  #1}}}
\newcommand{\sz}[1]{{{\color{black}  #1}}}

\newcommand{\bbb}{{\boldsymbol \beta}}

\newcommand{\ttt}{\boldsymbol \theta}

\newcommand{\YY}{\mbox{$\mathbf Y$}}

\newcommand{\xx}{\mathbf x}
\newcommand{\yy}{\mathbf y}

\newcommand{\argmax}{\operatornamewithlimits{arg\,max}}

\newtheorem{definition}{Definition}

\title{A Latent Gaussian Process Model for Analyzing Intensive Longitudinal Data}
\author{Yunxiao Chen\\
 Department of Statistics, London School of Economics and Political Science\\
 Siliang Zhang\\
 Shanghai Center for Mathematical Sciences, Fudan University}
\date{}

\begin{document}
\maketitle

\doublespacing

\begin{abstract}

\sz{Intensive longitudinal studies} are becoming progressively more prevalent across many social science areas, especially in psychology.
\sz{New technologies like smartphones, fitness trackers, and the Internet of Things
make it much easier than in the past for data collection in intensive longitudinal studies,
providing an opportunity to look deep into the underlying characteristics of individuals under a high temporal resolution.}
\yc{
In this paper, we introduce a new modeling framework for latent curve analysis that is more suitable for the analysis of intensive longitudinal data than existing latent curve models. Specifically, through the modeling of an individual-specific continuous-time latent process, some unique features of intensive longitudinal data are better captured, including intensive measurements in time and unequally spaced time points of observations.
Technically, the continuous-time latent process is modeled by a Gaussian process model.
This model can be regarded as a semi-parametric extension of the classical latent curve models and falls under the framework of structural equation modeling.
Procedures for parameter estimation and statistical inference are provided under an empirical Bayes framework
and evaluated by simulation studies.
We illustrate the use of the proposed model though the analysis of an ecological momentary assessment dataset.

}

\end{abstract}	

\section{Introduction}

\yc{Intensive longitudinal data are becoming progressively more prevalent across many social science areas, especially in psychology, catalysed by technological advances \citep[e.g., Chapter 1,][]{bolger2013intensive}.  Such data usually involve many repeated measurements that reflect individual-specific change process in high resolution, enabling researchers to answer deeper research questions of human behavioral patterns.
Due to the complex structure of intensive longitudinal data, statistical models play an important role in the analysis of such data.
}



\begin{figure}
  \centering
  \includegraphics[scale=2]{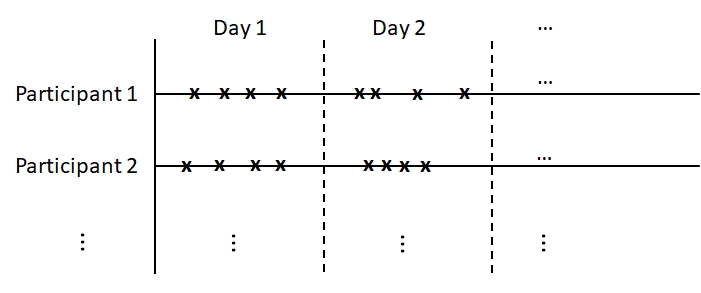}
  \caption{An illustration of the signal-contingent sampling scheme
  of an ecological momentary assessment.}\label{fig:design}
\end{figure}

\yc{In an intensive longitudinal study, repeated measurements are made intensively over time. Such data may involve
(1) a large number of time points, (2) individually-varying numbers of observations, (3) unequally spaced time points of observations, and (4) response data of various types (e.g., continuous, ordinal, etc.). For example, consider intensive longitudinal data from
ecological momentary assessment (EMA) under a signal-contingent sampling scheme \citep[see Chapter 5,][]{Conner2012getting}, which  repeatedly measures individuals' current behaviors and experiences in real time, in the individuals' natural environments.
Under this sampling scheme, participants are  ``beeped" at several (random) times a day to complete an electronic diary record  on psychological variables, such as symptoms or well-being. The assessments can last for many days (e.g. a month).
Such a design has been used to study, for example, borderline personality disorder \citep{trull2008affective}, adolescent smoking \citep{hedeker2012modeling}, and others.
We visualize this design in Figure~\ref{fig:design}, where the measurements happen at time points marked by ``{\bf x}". Under such a design, each individual may receive hundreds of repeated measurements at irregularly spaced time points. 
Depending on the measurement scale, one or multiple indicators may be recorded at each observation time point and the indicators can be either continuous or categorical. }

\yc{Latent curve models \citep[e.g.][]{bollen2006latent,duncan2013introduction, ram2015growth}, also known as latent growth models or growth curve models, are an important family of psychometric models for the analysis of longitudinal measurements. These models characterize the growth or change in an individual through the modeling of an individual-specific time-varying latent trait, where the latent trait often has a substantive interpretation, such as a cognitive ability, a psychopathological trait, or subjective well-being. Such models are typically formulated under the structural equation modeling framework.
In these models, each individual $i$ is represented by a
latent curve
$\{\theta_i(t): t \geq 0 \}$, which represents a time-varying latent trait.
At a given observation time $t$, the individual's response to a single or multiple items is assumed to be driven by his/her current latent trait level $\theta_i(t)$.

The classical latent curve models are developed for non-intensive longitudinal data (typically less than 10 times of measurement). Therefore, they often make strong assumptions on the functional form of $\theta_{i}(t)$.
For example, a linear latent curve model assumes that
$\theta_i(t) =  \beta_{i0} + \beta_{i1}t$, where $\beta_{i0}$ and $\beta_{i1}$ are the intercept and the slope of the curve, treated as individual specific latent variables. In other words, in this linear curve model, the latent curve
$\theta_i(t)$ is a random function, characterized by two random effects $\beta_{i0}$ and $\beta_{i1}$ that are often assumed to follow a bivariate normal distribution.
Although $\theta_i(t)$ can take slightly more complex forms (e.g., polynomial), the functional form of
$\theta_i(t)$ in the classical models is usually simple, which may not be suitable for analyzing individual change processes revealed by intensive longitudinal data, where the number of measurements may vary across different individuals.

To better capture the temporal pattern in intensive longitudinal data, more flexible latent curve models have been proposed
under the structural equation modeling framework. 
Depending on whether time is treated as discrete or continuous, these models can be classified into two categories. The discrete-time models are typically a hybrid of time series analysis models and the structural equation modeling framework. Specifically, the individual specific dynamic latent traits are modeled by a time series model, such as the autoregressive (AR) or vector autoregressive (VAR) models. Such models are usually known as the
latent variable-autoregressive latent trajectory models \citep{bianconcini2018latent} or dynamic structural equation models \citep{asparouhov2018dynamic}.
The continuous-time models typically assume that the dynamic latent traits follow a stochastic differential equation \citep[SDE;][]{oud2000continuous,voelkle2012sem,lu2015bayesian}. For example, \cite{lu2015bayesian} assume the dynamic latent trait to follow the Ornstein-Uhlenbeck Gaussian process \citep{uhlenbeck1930theory}, whose distribution is given by an SDE.

The above models have limitations. Discrete-time models may be over-simplified for intensive longitudinal data, for which measurement occurs in continuous time. In particular, when time points of measurements are irregularly spaced and different individuals have different numbers of measurements, it is difficult to organize intensive longitudinal data into the format of multivariate time-series data and then analyze using a discrete-time model. Arbitrarily transforming data into a multivariate time-series format
 is likely to introduce bias into the analysis, as
time lags between measurements, which may vary substantially among individuals, are ignored in the discrete-time formatting. 
In theory, these issues with discrete-time models can be addressed by taking a continuous-time model. However, existing continuous-time models are typically not straightforward to specify, estimate, and make inference upon, as latent stochastic differential equations are not straightforward to deal with either analytically or numerically. Moreover, limited by the form of stochastic differential equations, the existing continuous-time models for insensitive longitudinal data may not be rich enough.


In this paper, we propose new continuous-time latent curve models for the analysis of intensive longitudinal data that do not suffer from the issues with the existing models and better capture the unique features of intensive longitudinal data mentioned previously.
By imposing Gaussian process models \citep{rasmussen2006gaussian} on the latent curves $\{\theta_i(t): t \geq 0\}$, a general framework for latent curve modeling is developed. We call it the \textit{Latent Gaussian Process} (LGP) models.
In contrast to discrete-time models, the proposed models retain the flexibility of continuous-time models in dealing with  observations in a continuous time domain. 
In addition, this general framework contains models that are easier to specify and analyze than SDE-based models.

Technically, the proposed modeling framework can be viewed as a hybrid of
the latent Gaussian process model for functional data analysis \citep{hall2008modelling} and the  generalized multilevel structural equation modeling framework for longitudinal measurement \citep[e.g., Chapter 4,][]{skrondal2004generalized}. As will be shown in the sequel, many existing latent curve models, whether time is treated as continuous or discrete, can be viewed as special cases under the proposed general framework.
By making use of mathematical characterizations of Gaussian processes, methods for the parametrization of LGP models are provided. In addition, parameter estimation and statistical inference are carried out under an empirical Bayes framework, using
a Stochastic Expectation-Maximization (StEM) algorithm \citep{celeux1985sem, nielsen2000stochastic,zhang2018improved}.


The rest of the paper is organized as follows.
In Section~\ref{sec:model}, the classical latent curve models are reviewed under a unified framework of structural equation modeling
and then a new latent Gaussian process modeling framework is introduced that substantially generalizes the traditional models. 
The parametrization of latent Gaussian process models is discussed.  Estimation and statistical inference are discussed in Section~\ref{sec:inference}, followed by the computational details in Section~\ref{sec:estimation}.
Extension to the incorporation of covariates is discussed in Section~\ref{sec:covariates}. The proposed model is
evaluated in Section~\ref{sec:sim} through simulation studies and further
illustrated in Section~\ref{sec:real} via a real data example.
We end with concluding remarks in Section~\ref{sec:conclu}.

}

\section{Latent Gaussian Process Model}\label{sec:model}



\subsection{A Unified Framework for Latent Curve Analysis}\label{subsec:framework}

We first provide a unified framework for latent curve analysis.
We consider $N$ participants being measured longitudinally within a time interval $[0, T]$, where time is treated as continuous. For individual $i$, let $t_{is} \in [0, T]$ be the time that the $s$th measurement occurs
and $S_{i}$ be the total number of measurements received by individual $i$.
At each time $t = t_{i1}, ..., t_{iS_i}$, we observe a random vector
$\YY_i(t) = (Y_{i1}(t), ..., Y_{iJ}(t))^{\top}$, where
$Y_{ij}(t)$ can be either continuous or categorical, depending on the data type of the $j$th indicator.
In particular, 
the corresponding latent curve model is called a single-indicator model when $J = 1$ and
a multiple-indicator model when $J > 1$.
We denote $\yy_{i}(t) = (y_{i1}(t), ..., y_{iJ}(t))^\top$ as a realization of $\YY_i(t)$.
Moreover, each individual $i$ is associated with a latent curve $\theta_i(\cdot) = \{\theta_i(t): t \in [0, T] \}$, which can be regarded as a time-varying latent trait. Note that the above setting is quite general that includes
discrete-time longitudinal data as a special case, for which the observation time $t_{is}$ takes value in $\{0, 1, 2, ...\}$.


The latent curve model consists of two components: (1) a measurement model that specifies the conditional distribution of $\{\YY_i(t): t = t_{i1}, ..., t_{iS_i}\}$ given $\{\theta_{i}(t): t \in [0, T]\}$, and (2) a structural model that specifies the distribution of the random function $(\theta_{i}(t): t \in [0, T])$. 

\begin{figure}[h]
  \centering
  \includegraphics[scale=2]{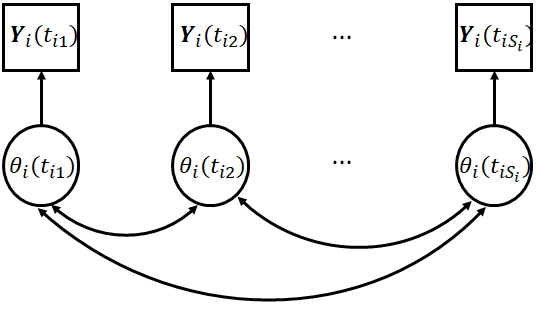}
  \caption{Path diagram for a unified latent curve model.}\label{fig:model}
\end{figure}

\paragraph{Measurement model.}
The measurement model assumes that the distribution of $\YY_i(t)$ only depends on $\theta_i(t)$, the latent trait level at the same time point,
but does not depend on the latent trait levels or responses at any other time points.
More precisely, it is assumed that
\begin{equation}\label{eq:cond}
f(\yy_{i}(t_{i1}), ..., \yy_{i}(t_{iS_i})\vert \theta_{i}(t), t\in [0, T])
= f(\yy_{i}(t_{i1}), ..., \yy_{i}(t_{iS_i})\vert \theta_{i}(t_{i1}), ..., \theta_{i}(t_{iS_i})),
\end{equation}
where $f(\yy_{i}(t_{i1}), ..., \yy_{i}(t_{iS_i})\vert \theta_{i}(t), t\in [0, T])$ denotes the probability density/mass function of the conditional distribution of $\YY_{i}(t_{i1})$, ..., $\YY_{i}(t_{iS_i})$  given the entire latent process $(\theta_i(t): t\in [0, T])$ and $f(\yy_{i}(t_{i1}), ..., \yy_{i}(t_{iS_i})\vert \theta_{i}(t_{i1}), ..., \theta_{i}(t_{iS_i}))$ denotes
the probability density/mass function of the conditional distribution of $\YY_{i}(t_{i1})$, ..., $\YY_{i}(t_{iS_i})$  given $\theta_{i}(t_{i1}), ..., \theta_{i}(t_{iS_i})$. Equation~\eqref{eq:cond} means that the latent trait level at any other time point is conditionally independent of the observed responses, given the latent trait levels at the corresponding time points of observation.
As visualized in Figure~\ref{fig:model}, it is further assumed that the conditional distribution~\eqref{eq:cond} has the following decomposition,
\begin{equation}\label{eq:cond2}
f(\yy_{i}(t_{i1}), ..., \yy_{i}(t_{iS_i})\vert \theta_{i}(t_{i1}), ..., \theta_{i}(t_{iS_i})) = \prod_{s= 1}^{S_i} g(\yy_i(t_{is}) \vert \theta_i(t_{is})),
\end{equation}
where $g(\yy_i(t) \vert \theta_i(t))$ is the conditional probability density/mass function of $\YY_i(t)$ given $\theta_i(t)$. The assumption in \eqref{eq:cond2} is conceptually similar to the widely used local independence assumption in latent variable models \citep[see Chapter 4,][]{skrondal2004generalized}.
%
Finally, we assume local independence among multiple indicators at each time $t$, i.e., $Y_{i1}(t)$, ..., $Y_{iJ}(t)$
are conditionally independent given $\theta_i(t)$. That is
\begin{equation}\label{eq:measurement}
g(\yy_i(t) \vert \theta_i(t)) = \prod_{j=1}^J g_j (y_{ij}(t) \vert \theta_i(t)),
\end{equation}
where $g_j (y_{ij}(t) \vert \theta_i(t))$ specifies the conditional distribution of the $j$th indicator
$Y_{ij}(t)$ given $\theta_i(t)$. The choice of $g_j$ depends on the type of the $j$th indicator. It is worth noting that the conditional distribution $g_j$ does not depend on time $t$, implying that the measurement is assumed to be time-invariant. 
Although commonly adopted in latent curve models \citep[e.g., Chapter 2,][]{bollen2006latent}, this assumption is quite strong and needs to be checked when applying such models to real data.


%

We provide several measurement model examples.
\begin{enumerate}
  \item Linear factor model for continuous response:
  \begin{equation}\label{eq:measurement1}
  Y_{ij}(t)\vert \theta_i(t) \sim N(a_j \theta_i(t)  + b_j, \sigma_j^2),
  \end{equation}
  where $a_j$, $b_j$, and $\sigma_j^2$ are model parameters. 

   \item Probit model for ordinal response ($Y_{ij} \in  \{0, 1, ..., n_j\}$):
  \begin{equation}\label{eq:PCR}
    P(Y_{ij}(t) = l \vert \theta_i(t)) = \Phi\left(b_{j,l+1}+a_j\theta_i(t)\right)-\Phi\left(b_{j,l}+a_j\theta_i(t)\right),
  \end{equation}
  where $$-\infty=b_{j,0} < b_{j,1} < b_{j,2} <\hdots<b_{j,n_j} < b_{j,n_j+1}=\infty.$$ $b_{j,l}$ and $a_{j}$ are model parameters, $l\in \{1, ..., n_j\}$ and $j = 1, ...,J$. When $n_j = 1$, $Y_{ij}$ degenerates to a binary response variable and the model \eqref{eq:PCR} becomes the well-known two-parameter normal-ogive model in item response theory \citep[Chapter 4,][]{embretson2000item}.

  Model \eqref{eq:PCR} can be specified alternatively through the introduction of latent responses. That is, define latent response
  \[
  Y^*_{ij}(t) = -a_j\theta_i(t) + \epsilon_{ij}(t)
  \]
  where $\epsilon_{ij}(t)$ is a noise term following a standard normal distribution. Then the observable response $Y_{ij}(t)$ can be viewed as a truncated version of $Y^*_{ij}(t)$, obtained by
  \[
  Y_{ij}(t) = l\ \text{if}\ b_{j,l} \leq Y^*_{ij}(t) < b_{j,l+1}.
  \]

\end{enumerate}
When the  multiple indicators contain a mixture of ordinal and continuous variables, the above models can be combined to model $Y_{i1}(t), ..., Y_{iJ}(t)$, since  the measurement models for different items can be specified independently given the local independence assumption.


\paragraph{Structural model.} The structural model specifies the distribution of the random function $\theta_{i}(t)$. We list a few examples below and refer the readers to \cite{bollen2006latent} for a comprehensive review.
\begin{enumerate}
  \item Linear trajectory model:
   \begin{equation}\label{eq:linear}
   \theta_{i}(t) = \beta_{i0} + \beta_{i1}t,
   \end{equation}
   where $\bbb_i = (\beta_{i0}, \beta_{i1})$ are individual specific random effects, following a bivariate normal distribution.
 \item Quadratic trajectory model:
    \begin{equation}\label{eq:quad}
     \theta_{i}(t) = \beta_{i0} + \beta_{i1}t + \beta_{i2}t^2,
    \end{equation}
   where $\bbb_i = (\beta_{i0}, \beta_{i1}, \beta_{i2})$ are individual specific random effects, following a trivariate normal distribution.
 \item Exponential trajectory model:
     \begin{equation}\label{eq:expo}
      \theta_{i}(t) = \beta_{i0} + \beta_{i1}\exp{(\gamma t)},
     \end{equation}
    where $\bbb_i = (\beta_{i0}, \beta_{i1})$ are individual specific random effects, following a bivariate normal distribution and $\gamma$ is a fixed effect parameter. 

\end{enumerate}

These models assume a simple functional form for $\theta_{i}(t)$. In particular, the realizations of $\theta_{i}(t)$ are restricted to linear, quadratic, and exponential functions for models~\eqref{eq:linear}-\eqref{eq:expo}, respectively.
Such models tend to be effective for non-intensive longitudinal data (typically
less than 10 measurements), but may not be flexible enough when having intensive longitudinal measurements which provide information in a high temporal resolution. In the rest of the paper, a general modeling framework is proposed, based on which more flexible structural models can be constructed.

\subsection{Gaussian Process Structural Model}

In what follows, we introduce a new framework for modeling $\theta_i(t)$ as a continuous-time stochastic process.
A key component of this framework is the Gaussian process model.

\begin{definition}[Gaussian Process]
A time continuous stochastic process $X(t)$ on time interval $[0, T]$ is a Gaussian process
if and only if
for every finite set of time points $t_1, ..., t_S \in [0, T]$,
$(X(t_1), ..., X({t_S}))$ is multivariate normal.
\end{definition}

We remark that a Gaussian process can be defined more generally on a real line. In this paper, we focus on Gaussian process on a bounded interval $[0, T]$, since real longitudinal data are collected within a certain time window.
Many widely used stochastic processes, including the Brownian motion, the Brownian bridge, and the Ornstein-Uhlenbeck process, are special cases of Gaussian process. Thanks to the flexibility,
nonlinearity, and inherent nonparametric structure, Gaussian processes have been widely used as a model for random functions for solving regression, classification, and dimension reduction problems \citep[Chapter 4,][]{rasmussen2006gaussian}.

Thanks to the normality, a Gaussian process is completely characterized by two components: (1) a mean function $m(t) = EX(t)$,  and (2) a kernel function $K(t, t')$ for the covariance structure, where $K(t, t') = Cov(X(t), X(t'))$. We provide a definition of a kernel function below.
\begin{definition}[Kernel Function]\label{def:kernel}
A bivariate function $K(t, t')$ is called a kernel function if
for every finite set of points $t_1, ..., t_S$, the matrix
$(K(t_i, t_j): i,j = 1, ..., S)$ is positive semidefinite.
\end{definition}

\begin{figure}
  \centering
  \includegraphics[scale=0.6]{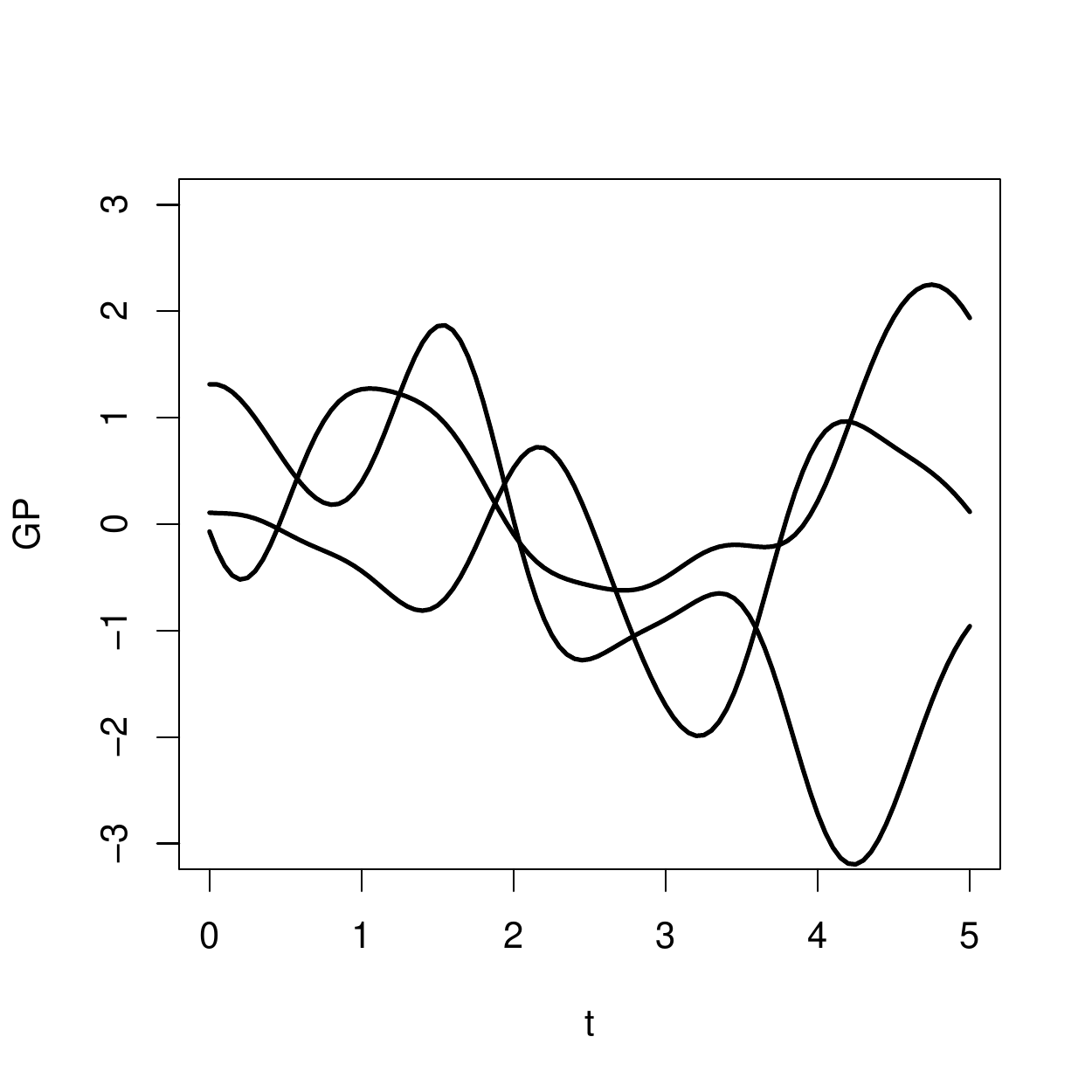}
  \caption{Sample paths from a Gaussian process, where $m(t) = 0$ and $K(t, t') = \exp\left(-\frac{(t-t')^2}{2 \times 0.5^2}\right)$}\label{fig:gp}
\end{figure}

Note that since $K(t, t') = Cov(X(t), X(t'))$, the matrix $(K(t_i, t_j): i,j = 1, ..., S)$ has to be positive semidefinite, because
it is the covariance matrix of $(X(t_1), ..., X({t_S}))$. On the other hand, it can be shown that for any kernel function $K$, there exists a Gaussian process whose covariance structure is given by the kernel \citep[Chapter 4,][]{rasmussen2006gaussian}. As an illustrative example,
Figure~\ref{fig:gp} shows three independent realizations from a Gaussian process,
with a mean function $m(t) = 0$ and a squared exponential kernel function $K(t,t') = \exp(-(t-t')^2/(2 \times 0.5^2))$.

\begin{definition}[Gaussian Process Structural Model]\label{def:gps}
We say the structural component of a latent curve model follows a Gaussian process structural model,
if $\{\theta_i(t): t\in [0, T]\}$ are independent and identically distributed (i.i.d.) Gaussian processes for $i = 1, ..., N$.
\end{definition}
We remark that the Gaussian process structural model assumption in Definition~\ref{def:gps} can be viewed as an extension of a commonly adopted assumption in unidimensional or multidimensional item response theory models where
individual-specific latent trait or traits are assumed to be i.i.d.
 univariate or multivariate normal. The difference is that, rather than having a random variable or random vector for each individual, each individual in the proposed model is characterized by a random function,
whose distribution is less straightforward to parameterize.


Combining a Gaussian process structural model and a measurement model as defined in Section~\ref{subsec:framework}, we obtain an LGP model. We point out that the examples \eqref{eq:linear}-\eqref{eq:expo} are all special cases of the LGP model. This is because, due to the multivariate normality of the random effects, for every finite set of time points $t_1, ..., t_S$,
$(\theta_{i}(t_1), ..., \theta_i(t_S))$ is multivariate normal. In addition, all the SDE based continuous-time latent curve
models also fall into this framework, when the noise component of the SDE is assumed to be Gaussian.
For example,  \cite{lu2015bayesian} assume the dynamic latent trait to follow the Ornstein-Uhlenbeck process \citep{uhlenbeck1930theory}. This process is a Gaussian process described by a stochastic differential equation with Gaussian noise. Furthermore, when the latent variables are assumed to be jointly normal,
the latent variable-autoregressive latent trajectory models \citep[see][]{bianconcini2018latent}, which are discrete-time
models,  can also be viewed as special cases under the current framework.

A Gaussian process is specified by a mean function $m(t)$ and a kernel function $K(t, t')$, whose choices should be problem specific. We denote the distribution of such a stochastic process by $\text{GP}(m,K)$.
In what follows, we discuss the parametrization of Gaussian process structural models.

\subsection{Parametrization of Gaussian Process Structural Model} \label{subsec:para}

Following the above discussion, we see that $\theta_i(t) = m(t) + \bar{\theta}_i(t)$, where $\bar{\theta}_i(t)$ is Gaussian process with mean 0 and kernel $K(t, t')$. This allows us to discuss the modeling of $m(t)$ and $K(t, t')$ separately, while in the classical latent curve models (e.g., \eqref{eq:linear}-\eqref{eq:expo}) the mean and kernel are modeled simultaneously. In particular, the mean process $m(t)$ can be viewed as the mean of $\theta_i(t)$, for individuals from a population of interest. Therefore, the mean function captures the mean level of the time-varying latent trait, possibly reflecting the trend and the periodicity of the dynamic latent trait at the population level. In addition,
the mean zero Gaussian process $\bar{\theta}_i(t)$ can be viewed as the  deviation from the mean process that is specific to individual $i$.

\paragraph{Mean function.} We consider the parametrization of the mean function $m(t)$, which is typically assumed to have  certain level of smoothness. 
Specifically, for the linear, the quadratic, and the exponential trajectory models mentioned in Section~\ref{subsec:framework},
the mean functions $m(t)$ take linear, quadratic, and exponential forms.

Under the current framework, $m(t)$ can be parameterized more flexibly. 
Specifically, we adopt a parametrization of $m(t)$ using basis functions. That is,
\begin{equation}\label{eq:meanfun}
m(t) = \alpha_0 + \alpha_1 b_1(t) + \cdots + \alpha_D b_D(t),
\end{equation}
where $b_1(t)$, ..., $b_D(t)$ are pre-specified basis functions on $[0, T]$. For example, when polynomial basis functions are used, $b_d(t) = t^{d}$, $d = 1, 2, ..., D$, where $D$ is the degree of the polynomial function. When cubic spline basis functions are used, $b_1(t) = t$, $b_2(t) = t^2$, $b_3(t) = t^3$, and $b_{3+d} = (t-\xi_d)^3_+$, where $d= 1, ..., D-3$, $\xi_d$ is the $d$th spline knot that is pre-specified on $[0, T]$, and $(t-\xi_d)^3_+ = (t-\xi_d)^3$ when $t > \xi_d$ and 0 otherwise. Alternative basis functions may also be used, such as Fourier basis, wavelets, and other spline basis functions. We refer the readers to  Chapter 3, \cite{ramsay1997functional} for a review of different basis functions.
We remark that the number of basis functions $D$ and the choices of basis function may be determined by data through model comparison.

 We remark  that if the dynamic trait $\theta_i(t)$ is assume to be a stationary process (i.e., the joint distribution of $\theta_i(t)$ does not change when the process is shifted in time), then the mean function does not depend on time $t$. In that case, the mean function can only have an intercept parameter, $m(t) = \alpha_0$.

\paragraph{Parameterizing kernel function.} One way to model the mean zero Gaussian process $\bar \theta_i (t)$ is by directly parameterizing the kernel function. In fact, different parametric kernel functions are available in the literature. We refer the readers to \cite[Chapter 4,][]{rasmussen2006gaussian} for a review. In what follows, we provide a few examples of kernel functions,  with a focus on kernels that lead to stationary mean zero Gaussian processes.
For such a kernel function $K(t, t')$, the value of $K(t, t')$ only depends on
the time lag $|t-t'|$, not the specific values of $t$ and $t'$. A stationary kernel should be used if the distribution of $\bar \theta_i (t)$ is believed to be invariant when the process is shifted in time.

\begin{enumerate}
  \item Squared exponential (SE) kernel:
  \begin{equation}\label{eq:sekernel}
K(t, t') =  c^2\exp\left(-\frac{(t-t')^2}{2\kappa^2}\right),
\end{equation}
where $c > 0$ and $\kappa > 0$ are two model parameters, known as the scale and the length scale parameters, respectively.
\item Exponential kernel:
  \begin{equation}\label{eq:expkernel}
K(t, t') =  c^2\exp\left(-\frac{|t-t'|}{2\kappa^2}\right),
\end{equation}
where $c > 0$ and $\kappa > 0$ are two model parameters that play similar roles as the ones in the SE kernel above.
\item Periodic kernel \citep{mackay1998introduction}:
 \begin{equation}\label{eq:perkernel}
K(t, t') =  c^2\exp\left(-\frac{2\sin^2(\pi|t-t'|/p)}{\kappa^2}\right),
\end{equation}
where $c > 0$ and $\kappa > 0$ are two model parameters that play similar roles as the ones in the two kernels above and $p$ is known as the period parameter which determines the periodicity of the kernel function.
\end{enumerate}

Mean zero Gaussian processes with different kernel functions have different properties. For example,
the mean zero Gaussian processes with an SE kernel tend to have smooth pathes. In fact,
a mean zero Gaussian process with the SE kernel is classified as one of the most smooth stochastic processes, according to the notion of mean square differentiability \citep[Chapter 1,][]{adler2010geometry}, a classical quantification of the smoothness of stochastic processes. This kernel function is widely used in statistical applications of Gaussian process. It will be further discussed in the sequel and be used in the data analysis. 

An alternative way of parameterizing the kernel is by directly modeling the mean zero Gaussian process, which can be done by using a linear basis function model.
Specifically, let $\phi_1(t)$, ..., $\phi_H(t)$ be $H$ pre-specified basis functions on $[0, T]$, such as spline basis, Fourier basis, or wavelet basis functions. The theory of functional principal component analysis provides an idea on
choosing better basis functions \citep[e.g.,][]{hall2008modelling}. 
Given the basis functions, the linear basis function model assumes that
\begin{equation}\label{eq:linear2}
\bar{\theta}_i(t) = \sum_{h=1}^H \omega_h Z_{ih} \phi_h(t),
\end{equation}
where $\omega_h$, $h = 1, ..., H$, are model parameters and $Z_{ih}$, $h = 1, ..., H$, are i.i.d. standard normal random variables. The model \eqref{eq:linear2} yields
$$K(t, t') = \sum_{h=1}^H \omega_h^2 \phi_h(t)\phi_h(t').$$
For finite $H$, this parametrization approach typically leads to a non-stationary kernel function.
Making use of the theory of reproducing kernel Hilbert space, essentially any mean zero Gaussian process can be approximated by the form of \eqref{eq:linear2} for sufficiently large $H$.

\begin{figure}
  \centering
  \includegraphics[scale=0.6]{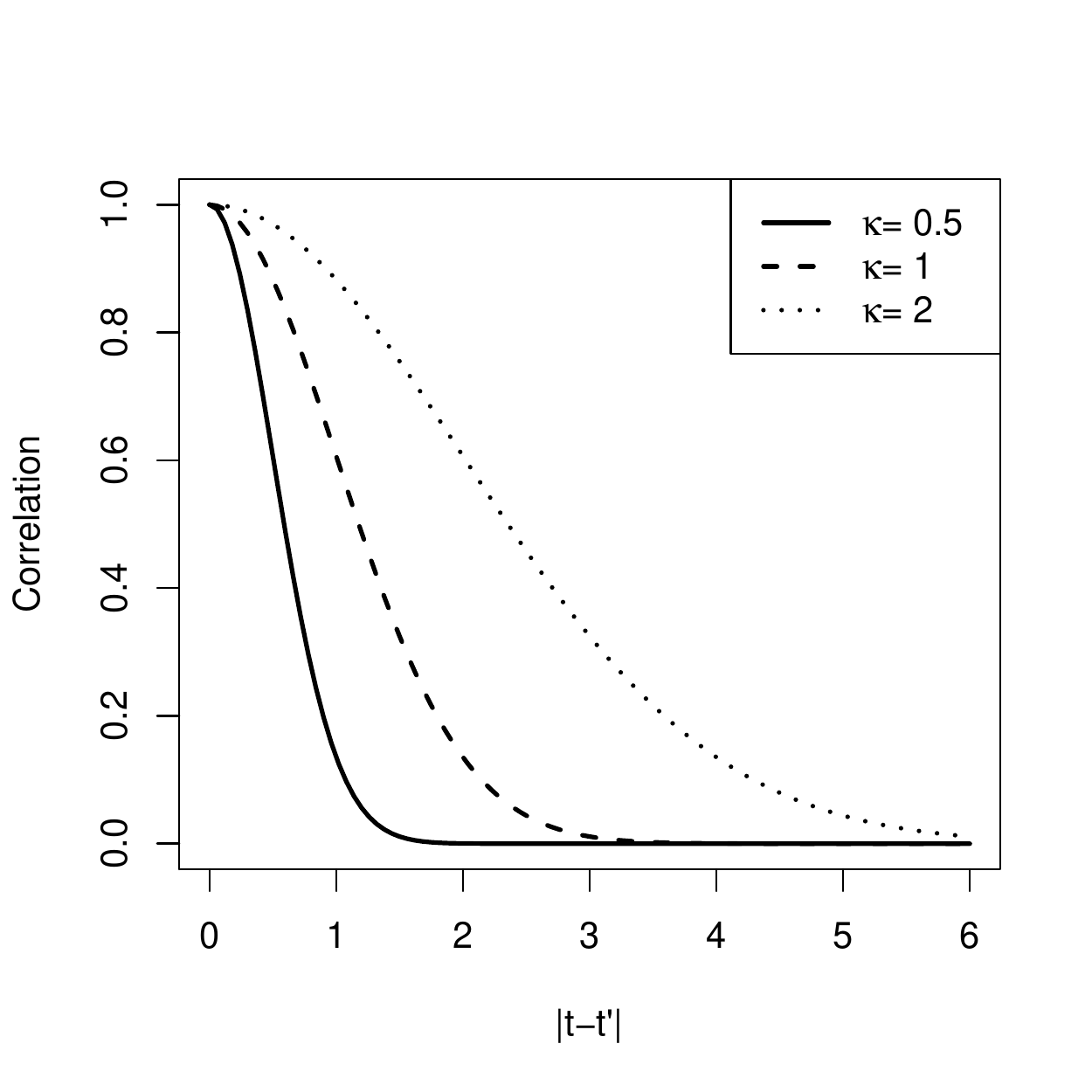}
  \caption{An illustration of the squared exponential kernel.}\label{fig:kernel}
\end{figure}

\begin{figure}
  \centering
  \includegraphics[scale=0.5]{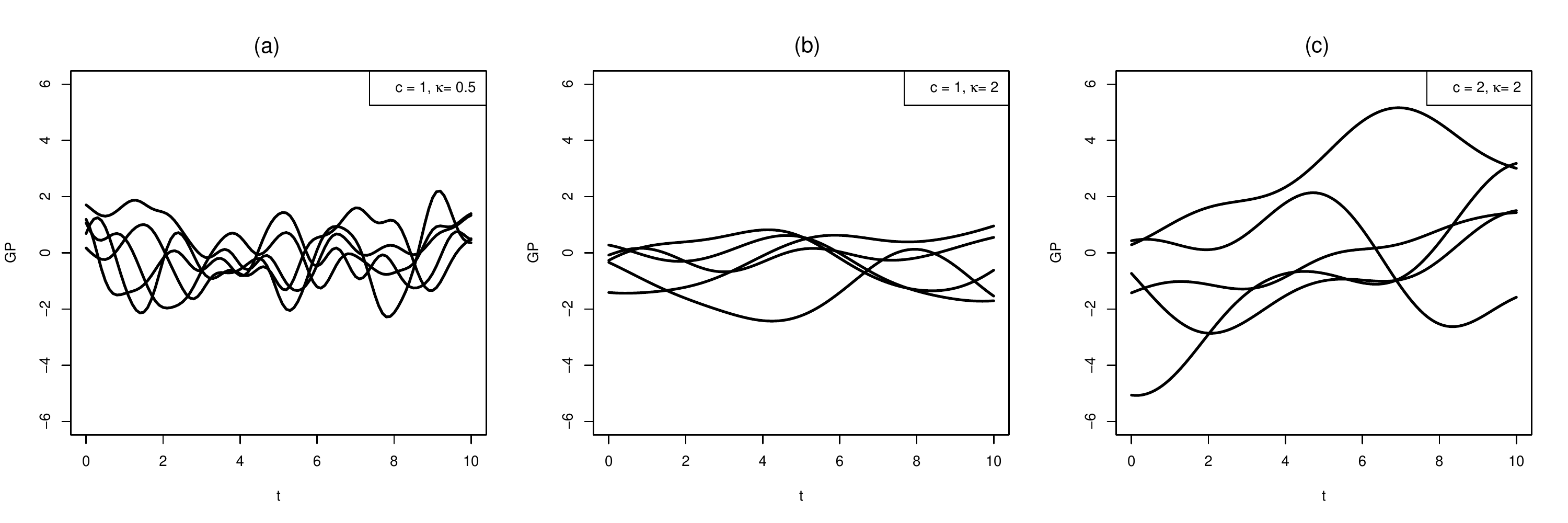}
  \caption{Sample paths from three Gaussian processes with mean 0 and SE kernels. The SE kernels differ by their values of the $c$ and the $\kappa$ parameters.}\label{fig:GP2}
\end{figure}

\paragraph{Squared exponential kernel.}
We further discuss on the properties of the SE kernel.
According to \eqref{eq:sekernel}, $Var(\theta_{i}(t)) = K(t,t) = c^2$.
\textit{The scale parameter $c$ thus captures the overall variation of the Gaussian process in the long run. }
Moreover, \textit{the length-scale parameter $\kappa$ captures the short-term temporal dependence}.
More precisely,
the correlation between $\theta_i(t)$ and $\theta_i(t')$ is given by
$$Cor(\theta_i(t), \theta_i(t')) = \frac{Cov(\theta_i(t), \theta_i(t'))}{\sqrt{(Var(\theta_i(t))\times Var(\theta_i(t')))}}= \exp\left(-\frac{(t-t')^2}{2\kappa^2}\right).$$
As shown in Figure~\ref{fig:kernel}, for each value of $\kappa$, the correlation decays towards zero as the time lag increases. The decaying rate is determined by the value of $\kappa$.
In particular, when the time lag $|t-t'| > 2\kappa$, the correlation is smaller than $\exp(-2) = 0.14$.
Moreover, for a given time lag, a smaller value of $\kappa$ implies a smaller correlation. Figure~\ref{fig:GP2} shows sample paths from three Gaussian processes with mean zero and SE kernels. Specifically, in panel (a), $c= 1, \kappa = 0.5$, in (b), $c = 1, \kappa = 2$, and in (c), $c = 2, \kappa = 2$. Panels (a) and (b) only differ by the values of the $\kappa$ parameter and the paths in panel (a) are from a Gaussian process with a smaller value of $\kappa$. The paths in panel (a) are more wiggly (i.e., have more short-term variation) than those in panel (b), since the Gaussian process in panel (a) has less temporal dependence. Panels (b) and (c) only differ by the values of $c$, due to which the paths in panel (c) have more variation in the long run.



\paragraph{Identifiability of the model parameters} 
Like many other structural equation models, constraints are needed to ensure model identifiability.
In particular, two constraints are needed, one to fix the scale of the latent process
and the other to avoid mean shift.  For instance, we consider a model combining
the mean function \eqref{eq:meanfun}, the measurement model \eqref{eq:measurement1}, and the SE kernel \eqref{eq:sekernel}.
To fix the scale in this model, we can either fix the scale parameter $c = 1$ in \eqref{eq:sekernel} or the first loading parameter $a_1 = 1$ in \eqref{eq:measurement}.
In addition, to avoid mean shift,
we can set either $\alpha_0 = 0$ in \eqref{eq:meanfun} or $b_1=0$ in \eqref{eq:measurement1}.


%
%
%



\section{Inference under LGP Model}\label{sec:inference}

The statistical inference under the proposed model can be classified into two levels, the population level and individual level. Both levels of inference may be of interest in the latent curve analysis.
The population level inference considers the estimation of the parameters in both the measurement and structural models. The individual level inference focuses on the posterior distribution of
$\theta_i(t)$ given data from each individual $i$ when the measurement and the structural models are known (e.g. obtained from the population level inference).

\paragraph{Population level inference.} We use $\Psi$ to denote all the model parameters, including parameters from both the measurement and structural models.  As mentioned above, constraints may be imposed on $\Psi$ to ensure model identifiability. Our likelihood function can be written as
\begin{equation}\label{eq:likelihood}
L(\Psi) = \prod_{i=1}^N \int \prod_{s=1}^{S_i} \prod_{j=1}^J g_j (y_{ij}(t_{is}) \vert \theta_{is}) f_i(\theta_{i1}, ..., \theta_{iS_i}) d\theta_{i1} ... d\theta_{iS_i},
\end{equation}
where $f_i(\theta_{i1}, ..., \theta_{iS_i})$ is the density function of an $S_i$-variate
normal distribution with mean
$(m(t_{i1}), \cdots m(t_{iS_i}))$ and covariance matrix $(K(t,t'): t, t' = t_{i1}, ..., t_{iS_i})$. Note that this likelihood function is the marginal likelihood of data in which the latent curves are integrated out. The maximum likelihood estimator of $\Psi$ is defined as $\hat \Psi = \arg\max_{\Psi} L(\Psi)$, whose computation is discussed in Section~\ref{sec:estimation}.
We then obtain the estimated mean and kernel functions by plugging in $\hat \Psi$.


\paragraph{Individual level inference.}
Similar to the classical latent curve analysis, the current modeling framework also allows for statistical inference on the latent curve of each individual. For ease of exposition, we assume both the measurement and the structural models are known when making individual level inference. In practice, we can first estimate the model parameters and then treat the estimated model as the true one in making the individual level inference.
For individual $i$, whether or not measurement occurs at time $t^*$, one can infer on $\theta_i(t^*)$ based on the posterior distribution of $\theta_i(t^*)$ given $\yy_{i}(t_{i1}), ..., \yy_{i}(t_{iS_i})$.
By sweeping $t^*$ over the entire interval $[0, T]$, one obtains
the posterior mean of $\theta_i(t)$ as a function of $t$, which serves as a point estimate of individual $i$'s latent curve. When calculated under the estimated model, we call the posterior mean of $\theta_i(t)$ the Expected A Posteriori (EAP) estimate of individual $i$'s latent curve and denote it by $\hat \theta_i(t)$. It mimics the EAP estimate of an individual's latent trait level in item response theory \citep[e.g.][]{embretson2000item}.


\section{Computation}\label{sec:estimation}

In this section, we elaborate on the computational details.

\subsection{Individual Level Inference}\label{subsec:ind}

We first discuss computing the posterior distribution of $\theta_i(t^*)$ given $\yy_{i}(t_{i1}), ..., \yy_{i}(t_{iS_i})$, for any time $t^*$, when both the measurement and the structural models are given. We denote the density of this posterior distribution by $h(\theta \vert \yy_{i}(t_{i1}), ..., \yy_{i}(t_{iS_i}))$.
Following equation~\eqref{eq:cond} of the measurement model,
$\theta_i(t^*)$ and $(\YY_{i}(t_{i1}), ..., \YY_{i}(t_{iS_i}))$ are conditionally independent given $(\theta_i(t_{i1}), ..., \theta_i(t_{iS_i}))$. Consequently,
\begin{equation}\label{eq:posterior}
\begin{aligned}
&h(\theta \vert \yy_{i}(t_{i1}), ..., \yy_{i}(t_{iS_i})) \\
=& \int h_1(\theta\vert \theta_1, ..., \theta_{S_i})
h_2(\theta_1, ..., \theta_{S_i} \vert \yy_{i}(t_{i1}), ..., \yy_{i}(t_{iS_i})) d\theta_1...d\theta_{S_i},
\end{aligned}
\end{equation}
where $h_1(\theta\vert \theta_1, ..., \theta_{S_i})$ denotes the conditional distribution of $\theta_i(t^*)$ given
$(\theta_i(t_{i1}), ..., \theta_i(t_{iS_i}))$ and $h_2(\theta_1, ..., \theta_{S_i} \vert \yy_{i}(t_{i1}), ..., \yy_{i}(t_{iS_i}))$ denotes the posterior distribution of $(\theta_i(t_{i1}), ..., \theta_i(t_{iS_i}))$ given the observed responses. 
Specifically, since $(\theta_i(t^*), \theta_i(t_{i1}), ..., \theta_i(t_{iS_i})$ follows a multivariate normal distribution with mean $(m(t^*), m(t_{i1}), ..., m(t_{iS_i}))$ and covariance matrix $(K(t,t'): t, t' = t^*, t_{i1}, ..., t_{iS_i})$, $h_1(\theta\vert \theta_1, ..., \theta_{S_i})$ is still normal,
for which the
mean $\mu(\theta_1, ..., \theta_{S_i})$ and variance $\sigma^2(\theta_1, ..., \theta_{S_i})$ have analytic forms.
Specifically,
$$\mu(\theta_1, ..., \theta_{S_i}) = m(t^*) + \Sigma_{12}\Sigma_{22}^{-1}(\ttt - \boldsymbol\mu) \mbox{~~and~~} \sigma^2(\theta_1, ..., \theta_{S_i}) = K(t^*, t^*) - \Sigma_{12}\Sigma_{22}^{-1}\Sigma_{21},$$
where $\ttt = (\theta_1, ..., \theta_{S_i})^\top$, $\boldsymbol\mu = (m(t_{i1}), ..., m(t_{iS_i}))^\top$, $\Sigma_{12} = (K(t^*, t_{i1}), ..., K(t^*, t_{iS_i}))$, $\Sigma_{22} = (K(t,t'): t, t' = t_{i1}, ..., t_{iS_i})$, and $\Sigma_{21} = \Sigma_{12}^\top$.
Then the posterior mean of $\theta_i(t^*)$ is given by
\begin{equation}\label{eq:postmean}
\begin{aligned}
\int \mu(\theta_1, ..., \theta_{S_i})
h_2(\theta_1, ..., \theta_{S_i} \vert \yy_{i}(t_{i1}), ..., \yy_{i}(t_{iS_i})) d\theta_1...d\theta_{S_i}.
\end{aligned}
\end{equation}
In addition, the $\alpha$-level quantile of the posterior distribution is given by
\begin{equation}\label{eq:quantile}
\begin{aligned}
\int (\mu(\theta_1, ..., \theta_{S_i}) + z_\alpha \sigma(\theta_1, ..., \theta_{S_i}))
h_2(\theta_1, ..., \theta_{S_i} \vert \yy_{i}(t_{i1}), ..., \yy_{i}(t_{iS_i})) d\theta_1...d\theta_{S_i},
\end{aligned}
\end{equation}
where $z_{\alpha}$ is the $\alpha$-level quantile of a standard normal distribution.

Under the linear factor model \eqref{eq:linear}, $(\theta_i(t^*), \theta_i(t_{i1}), ..., \theta_i(t_{iS_i}), \YY_i(t_{i1}), ..., \YY_i(t_{iS_i}))$ are jointly normal. Consequently, \eqref{eq:posterior}-\eqref{eq:quantile} have analytical forms. Under other measurement models, \eqref{eq:postmean} and \eqref{eq:quantile} can be approximated by using Monte Carlo samples
from the posterior distribution $h_2(\theta_1, ..., \theta_{S_i} \vert \yy_{i}(t_{i1}), ..., \yy_{i}(t_{iS_i}))$.
Specifically, let $(\theta_1^{(l)}, ..., \theta_{S_i}^{(l)}), l = 1, ..., L,$ be $L$ Monte Carlo samples.
Then we approximate the mean and $\alpha$-level quantile of the posterior distribution of $\theta_i(t^*)$
by
\begin{equation}\label{eq:apprpost}
\begin{aligned}
&\frac{1}{L}\sum_{l=1}^L \mu(\theta_1^{(l)}, ..., \theta_{S_i}^{(l)}),\\
&\frac{1}{L}\sum_{l=1}^L \mu(\theta_1^{(l)}, ..., \theta_{S_i}^{(l)}) + z_\alpha \sigma(\theta_1^{(l)}, ..., \theta_{S_i}^{(l)}).
\end{aligned}
\end{equation}
Markov chain Monte Carlo (MCMC) methods can be used to obain Monte Carlo samples from the posterior distribution $h_2(\theta_1, ..., \theta_{S_i} \vert \yy_{i}(t_{i1}), ..., \yy_{i}(t_{iS_i}))$. For example, a Gibbs sampler is developed that efficiently samples from this posterior distribution under the probit model \eqref{eq:PCR} for ordinal response data. This sampler, described as follows,  makes use of the latent response formulation of the  probit model  \eqref{eq:PCR}.

  \begin{itemize}
    \item[] \textbf{Step 1}: For $i=1, ...,N,$ $j = 1, ..., J$, $s = 1, ..., S_i$,  sample $y_{ij}^{*}(t_{is})$ from a truncated normal distribution
         that truncates a normal distribution $N(-a_j\tilde \theta_i(t_{is}), 1)$ by interval $[d_{j,y_{ij}(t_{is})}, d_{j,y_{ij}(t_{is})+1}]$, where $\tilde \theta_i(t_{is})$ is some initial value of $\theta_i(t_{is})$.

    \item[] \textbf{Step 2}: For $i=1, ...,N,$, given $y_{ij}^{*}(t_{is})$s, we update $(\tilde \theta_i(t_{i1}), ..., \tilde \theta_i(t_{iS_i}))$, by sampling from
    \[
    h_3(\theta_1, ...,\theta_{S_i}|\yy_{i}^{*}(t_{i1}), ...,\yy^{*}_{i}(t_{iS_i})),
    \]
    where $\yy_i^*(t) = (y_{i1}^{*}(t), ..., y_{iJ}^{*}(t))$ and $h_3$ denotes the conditional distribution of $(\theta_{i}(t_{i1}), ...,\theta_{i}(t_{iS_i}))$ given the ideal responses $\yy_{i}^{*}(t_{i1}), ...,\yy^{*}_{i}(t_{iS_i})$. It is worth noting that this conditional distribution is multivariate normal, because
     $\theta_{i}(t_{i1})$, ..., $\theta_{i}(t_{iS_i})$, $\yy_{i}^{*}(t_{i1})$, ..., $\yy^{*}_{i}(t_{iS_i})$ are jointly normal. The observed data $\yy_{i}(t_{i1})$, ..., $\yy_{i}(t_{iS_i})$ are not conditioned upon, because $\theta_{i}(t_{i1})$, ..., $\theta_{i}(t_{iS_i})$ are conditionally independent of the observed data when given the latent responses $\yy_{i}^{*}(t_{i1})$, ..., $\yy^{*}_{i}(t_{iS_i})$.
%
  \end{itemize}
We point out that both steps can be efficiently computed, because step 1 only involves sampling from univariate truncated normal distributions and step 2 only involves sampling from multivariate normal distributions. Well-developed samplers exist for both steps.


\subsection{Population Level Inference}\label{subsec:popinf}

We now discuss the computation for maximizing the likelihood function \eqref{eq:likelihood}. Under the linear factor model \eqref{eq:linear}, the Expectation-Maximization (EM) algorithm \citep{dempster1977maximum} is used to optimize \eqref{eq:likelihood}, where the E-step is in a closed form due to the joint normality of data and latent variables. The implementation of this EM algorithm is standard and thus we omit the details here.

Under other measurement models, the classical EM algorithm is typically computationally infeasible when the number of time points is large, in which case the E-step of the algorithm involves a high-dimensional integral that does not have an analytical form.
We adopt a stochastic EM (StEM) algorithm \citep{celeux1985sem,diebolt1996stochastic,zhang2018improved} which
avoids the numerical integration in the E-step of the standard EM algorithm \citep[]{dempster1977maximum,bock1981marginal} by Monte
Carlo simulations.
The convergence properties of the StEM algorithm
are established in \cite{nielsen2000stochastic}.
Similar to the EM algorithm, the StEM algorithm iterates between two steps, the StE step and the M step. Let $\Psi^{(0)}$ be the initial parameter values and
$(\tilde \theta_{i1}^{(0)}, \cdots, \tilde \theta_{iS_i}^{(0)}), i = 1, ..., N,$
be the initial values of person parameters. In each step $l$ ($l \geq 1$), the following StE step and M step are performed.

\begin{enumerate}
  \item[] \textbf{StE step}: For $i = 1, ..., N$,
  sample $(\tilde \theta_{i1}^{(l)}, \cdots, \tilde \theta_{iS_i}^{(l)})$ from $$h_2(\theta_1, ..., \theta_{S_i} \vert \yy_{i}(t_{i1}), ..., \yy_{i}(t_{iS_i}); \Psi^{(l-1)}),$$
  the conditional distribution of
  $(\theta_i{(t_{i1})}, \cdots, \theta_i{(t_{iS_i})})$ given $(\yy_{i}(t_{i1}), ..., \yy_{i}(t_{iS_i}))$
  under parameters $\Psi^{(l-1)}$. For the probit model \eqref{eq:PCR}, we use the Gibbs sampler described in Section~\ref{subsec:ind} to sample from $h_2(\theta_1, ..., \theta_{S_i} \vert \yy_{i}(t_{i1}), ..., \yy_{i}(t_{iS_i}); \Psi^{(l-1)})$.

  \item[] \textbf{M step}: Obtain parameter estimate
  \begin{equation}\label{eq:M-step}
  \Psi^{(l)} = \argmax_{\Psi} \sum_{i=1}^N l(\yy_{i}(t_{i1}), ..., \yy_{i}(t_{iS_i}), \tilde \theta_{i1}^{(l)}, \cdots, \tilde \theta_{iS_i}^{(l)}; \Psi),
  \end{equation}
  where
  \begin{equation}\label{eq:complete}
  \begin{aligned}
  &l(\yy_{i}(t_{i1}), ..., \yy_{i}(t_{iS_i}), \tilde \theta_{i1}^{(l)}, \cdots, \tilde \theta_{iS_i}^{(l)}; \Psi) \\
  =& \sum_{s= 1}^{S_i}\left[\sum_{j=1}^J \log g_j(y_{ij}(t_{is})\vert \tilde \theta_{is}^{(l)})\right] + \log f_i(\tilde \theta_{i1}^{(l)}, ..., \tilde \theta_{iS_i}^{(l)})
  \end{aligned}
  \end{equation}
  is the complete data log-likelihood of a single observation. Note that $g_j$ and $f_i$ are defined in \eqref{eq:measurement} and \eqref{eq:likelihood}, respectively, containing model parameters. In our implementation, the optimization is done using the  L-BFGS-B algorithm \citep{liu1989limited}.
\end{enumerate}
The final estimate of $\Psi$ is given by
the average of $\Psi^{(l)}$s from
the last $m$ iterations, i.e.,
\begin{equation}\label{eq:final}
\hat \Psi = \frac{1}{m} \sum_{l = m_0+1}^{m_0+m} \Psi^{(l)}.
\end{equation}
As shown in \cite{nielsen2000stochastic},
$\hat \Psi$ can approximate the maximum likelihood estimator sufficiently accurately,
when $m_0$ and $m$ are large enough.

\section{Incorporation of Covariates}\label{sec:covariates}

\begin{figure}[h]
  \centering
  \includegraphics[scale=1.5]{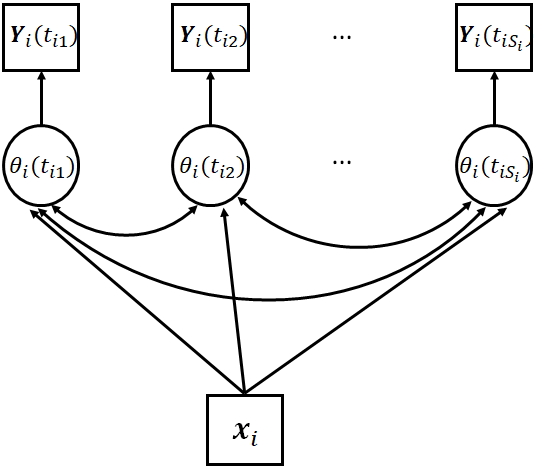}
  \caption{Path diagram of a latent curve model with covariates.}\label{fig:model2}
\end{figure}

In practice, individual specific covariates are often collected and incorporated into the latent curve analysis. As visualized in the path diagram in Figure~\ref{fig:model2}, covariates $\xx_i$ can be further added to the structural model to explain how the distribution of the latent curves depends on the covariates. A specific type of covariates of interest is group membership, such as experimental versus control and  female versus male. Latent curve analysis that incorporates discrete group membership as covariates in the structural model is referred to as the analysis of groups \citep[Chapter 6,][]{bollen2006latent}.

Covariates can be easily handled under the proposed framework. For example, when discrete group membership may affect the mean function of the latent curve, we let parameters in $m(t)$ to be group-specific. Similarly, we may also allow parameters in  $K(t, t')$ to depend on the group membership. Quantitative covariates, such as age, can also be incorporated into the current model.
The mean and kernel functions are denoted by $m_{\xx_i}(t)$ and $K_{\xx_i}(t,t')$ when they depend on the covariates. The tools for the inference and computation discussed above can be easily generalized.

\section{Simulation}\label{sec:sim}

The proposed modeling framework and the estimation procedures are further evaluated by simulation studies.

\subsection{Study I}

We first evaluate the parameter recovery using the EM algorithm, under a setting similar to the real data example in Section~\ref{sec:real}, except that  a single group is considered in this study.
In particular, it is assumed that each participant is measured for 25 consecutive days, with four measurements per day. Such a design results in 100 times of measurement. The time points of the four measurements are randomly sampled within a day.
And we consider a measurement model with a single indicator.
 More precisely, given the observation time,  the model is specified as follows.

\begin{equation*}
\begin{aligned}
Y_{i1}(t)\vert \theta_i(t) &\sim N(\theta_{i}(t), \sigma^2),\\
\theta_i(\cdot) &\sim GP(m, K),
\end{aligned}
\end{equation*}
where $m(t) = \alpha$ and $K(t, t') = c^2 \exp(-(t - t')^2/(2\kappa^2))$. The true model parameters are specified in Table~\ref{tab:sim1}. Two sample sizes are considered, including $N = 50$ and $N = 100$. The simulation under each sample size is repeated for 100 times, based on which the mean squared error (MSE) for parameter estimation is calculated. According to the MSE for parameter estimation presented Table~\ref{tab:sim1}, the parameter estimation is very accurate under the current simulation settings and the estimation accuracy improves as the sample size increases.

We further illustrate the performance of the individual level inference based on the $L^2$ distance between $\theta_i(t)$ and its EAP estimate $\hat \theta_{i}(t)$, where the distance is defined as
$$d_i = \sqrt{\int_0^T (\theta_i(t) - \hat \theta_i(t))^2 dt}.$$
In particular, $d_i$ quantifies the inaccuracy of estimating the latent curve $\theta_i(t)$ by $\hat \theta_i(t)$.
The $L^2$ distance between $\theta_i(t)$ and $\hat \alpha$,
$$e_i = \sqrt{\int_0^T (\theta_i(t) - \hat \alpha)^2 dt},$$
is used as a reference for $d_i$ that quantifies the inaccuracy of estimating $\theta_i(t)$ by the estimate of the population mean $\hat \alpha$. The ratio $d_i/e_i$ serves as a measure of inaccuracy in estimating the latent curve of individual $i$, in which the difficulty in estimating the curve has been taken into account by the denominator $e_i$. The smaller the ratio is, the more accurate the latent curve $\theta_i(t)$ is estimated in a relative sense (relative to the overall difficulty in estimating $\theta_i(t)$ measured by $e_i$).

\begin{figure}
\centering
\begin{subfigure}{.49\textwidth}
  \centering
  \includegraphics[width=1\linewidth]{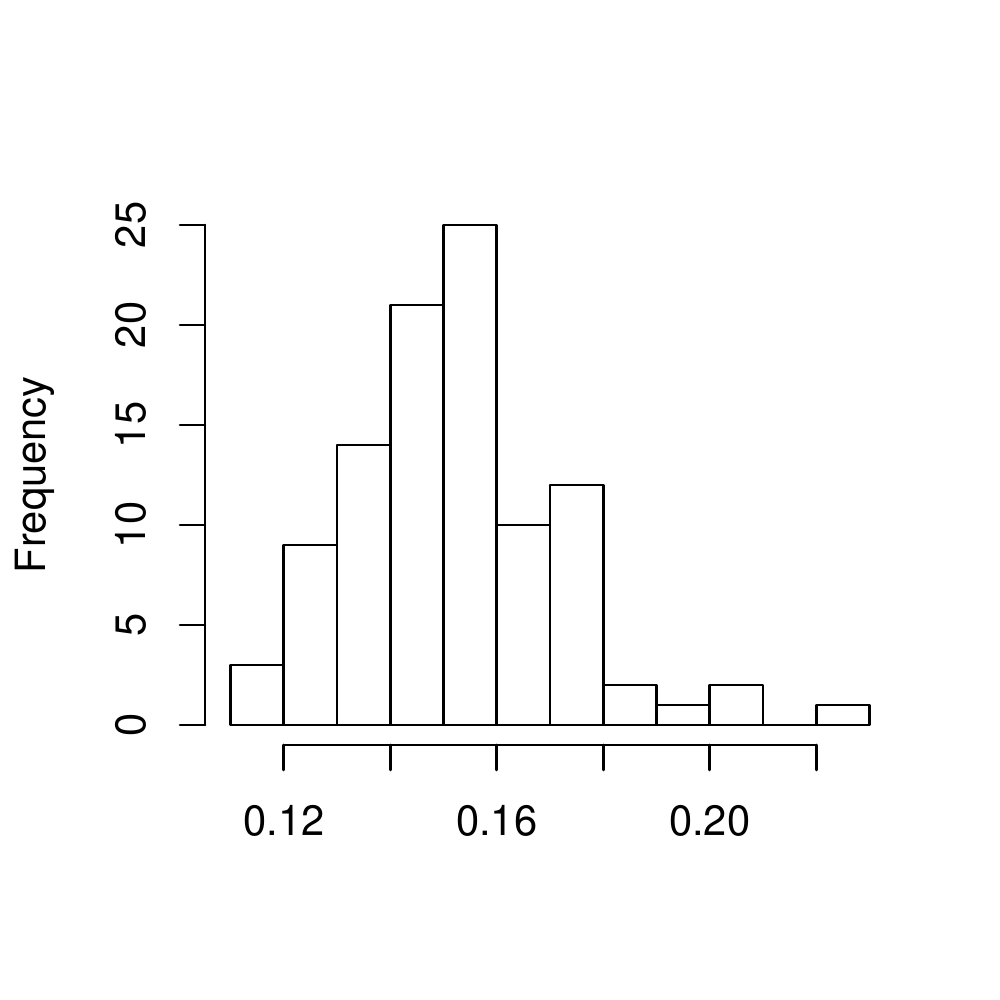}
  \caption{Histogram of the ratios $d_i$/$e_i$ for all individuals from a randomly selected dataset.}
\end{subfigure}%
\begin{subfigure}{.49\textwidth}
  \centering
  \includegraphics[width=1\linewidth]{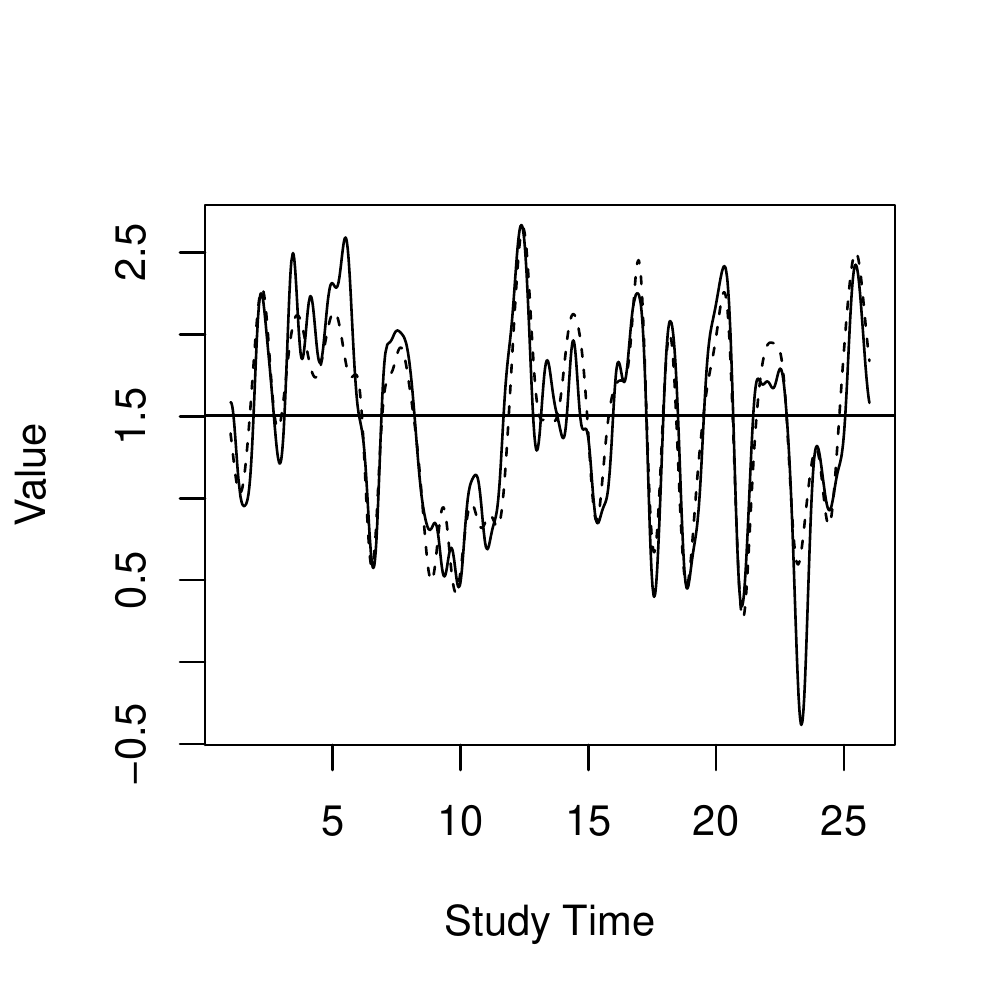}
  \caption{$\theta_i(t)$ (solid line) versus $\hat \theta_i(t)$ for an individual with $d_i/e_i = 0.14$}
\end{subfigure}
\\
\centering
\begin{subfigure}{.49\textwidth}
  \centering
  \includegraphics[width=1\linewidth]{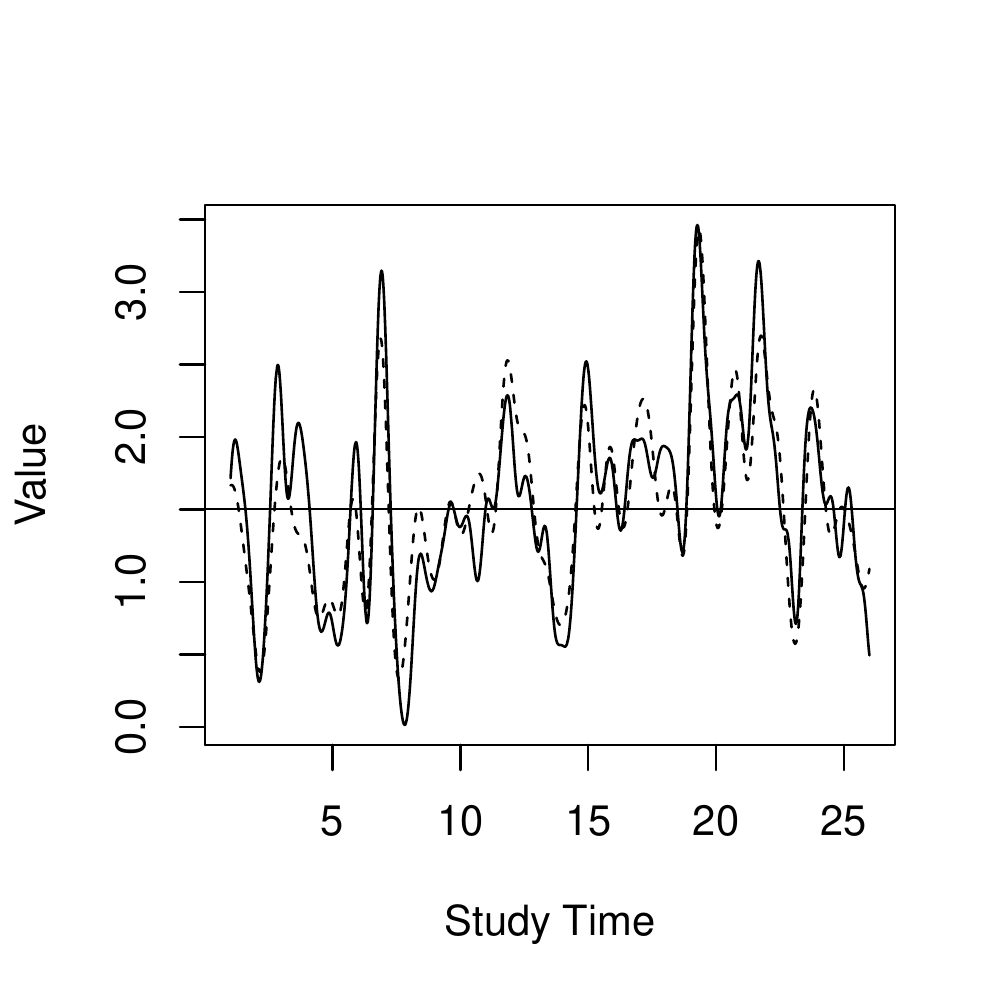}
  \caption{$\theta_i(t)$ (solid line) versus $\hat \theta_i(t)$ for an individual with $d_i/e_i = 0.15$}
\end{subfigure}
\begin{subfigure}{.49\textwidth}
  \centering
  \includegraphics[width=1\linewidth]{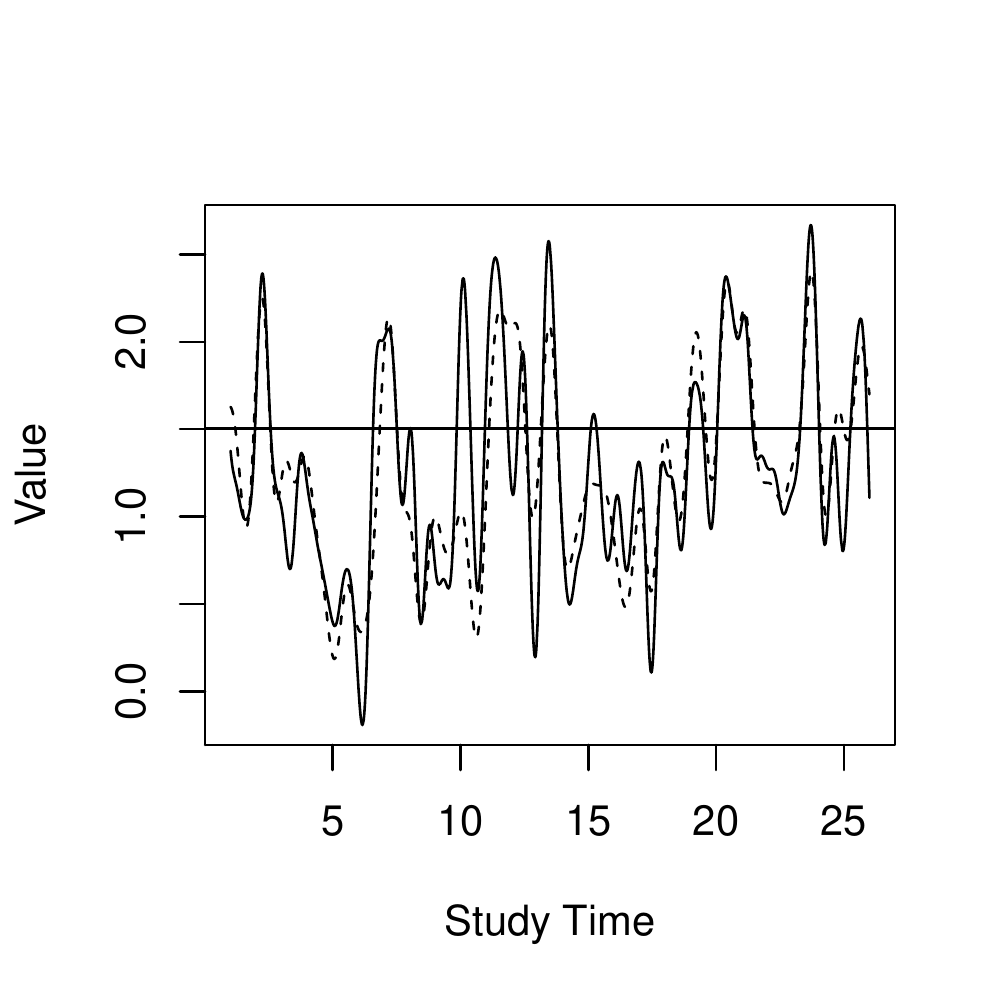}
  \caption{$\theta_i(t)$ (solid line) versus $\hat \theta_i(t)$ for an individual with $d_i/e_i = 0.20$}
\end{subfigure}
    \caption{Simulation Study I: Results on individual level inference.}\label{fig:11}
\end{figure}

In panel (a) of Figure~\ref{fig:11}, we show the histogram of the ratios $d_i/e_i$ for all individuals from a randomly selected dataset among all replications when the sample size $N = 100$. As we can see, $d_i$ is much smaller than $e_i$, implying that $\hat \theta_i(t)$ estimates $\theta_i(t)$ very accurately.  Panels (b)-(d) of Figure~\ref{fig:11} show $\theta_i(t)$, $\hat \theta_i(t)$, as well as $\hat \alpha$ for three randomly selected individuals from the same dataset. According to these plots,  the true latent curves  are well approximated by their EAP estimates.

\begin{table}
  \centering
  \begin{tabular}{lcccccc}
    \hline
    &$\alpha$ &$c^2$  &$\kappa$ &$\sigma^2$\\
    \hline
    True    &1.5&0.4&0.3&0.1\\
    \hline
    MSE($N = 50$)&$2.7 \times 10^{-4}$  &$2.2 \times 10^{-4}$  &$1.7 \times 10^{-4}$   &$1.3 \times 10^{-5}$\\
    MSE($N = 100$)& $1.2 \times 10^{-4}$ & $9.2 \times 10^{-5}$ & $1.5 \times 10^{-4}$ & $1.2\times 10^{-5}$\\
    \hline
  \end{tabular}
  \caption{Simulation Study I: Simulation results on the parameter recovery for an LGP model with a linear factor measurement model.}\label{tab:sim1}
\end{table}

\subsection{Study II}

We now consider a simulation study whose setting is the same as Study I except for a different measurement model component. In particular, we consider ordinal response data generated by the probit model \eqref{eq:PCR}.
Specifically,  the measurement at each time point is assumed to be based on five polytomous items, each with three ordinal categories (i.e., $n_j = 3$). The true model parameters are given in Table~\ref{tab:sim3}. Note that we fix $a_1 = 1$ and $d_{1,1} = 0$ in both the true model and the estimation procedure for model identifiability.



The simulation under each sample size is repeated for 100 times. For each simulated dataset, the model parameters are estimated using the stochastic EM algorithm described in Section~\ref{subsec:popinf}, based on a random initial value. The two tuning parameters $m_0$ and $m$ of the algorithm are set to be 100 and 200, respectively. The estimation accuracy measured by mean squared error is shown in Table~\ref{tab:sim3}, which indicates an accurate estimation result.
The running time of the stochastic EM algorithm for one dataset with $N = 100$ is around 10 minutes\footnote{The study is conducted on a personal computer with specifications: Processor 2.2 GHz Intel Core i7; Memory 8 GB 1600 MHz DDR3.}. It can be further speeded up by parallel computing.

\begin{figure}
\centering
\includegraphics[width=.5\linewidth]{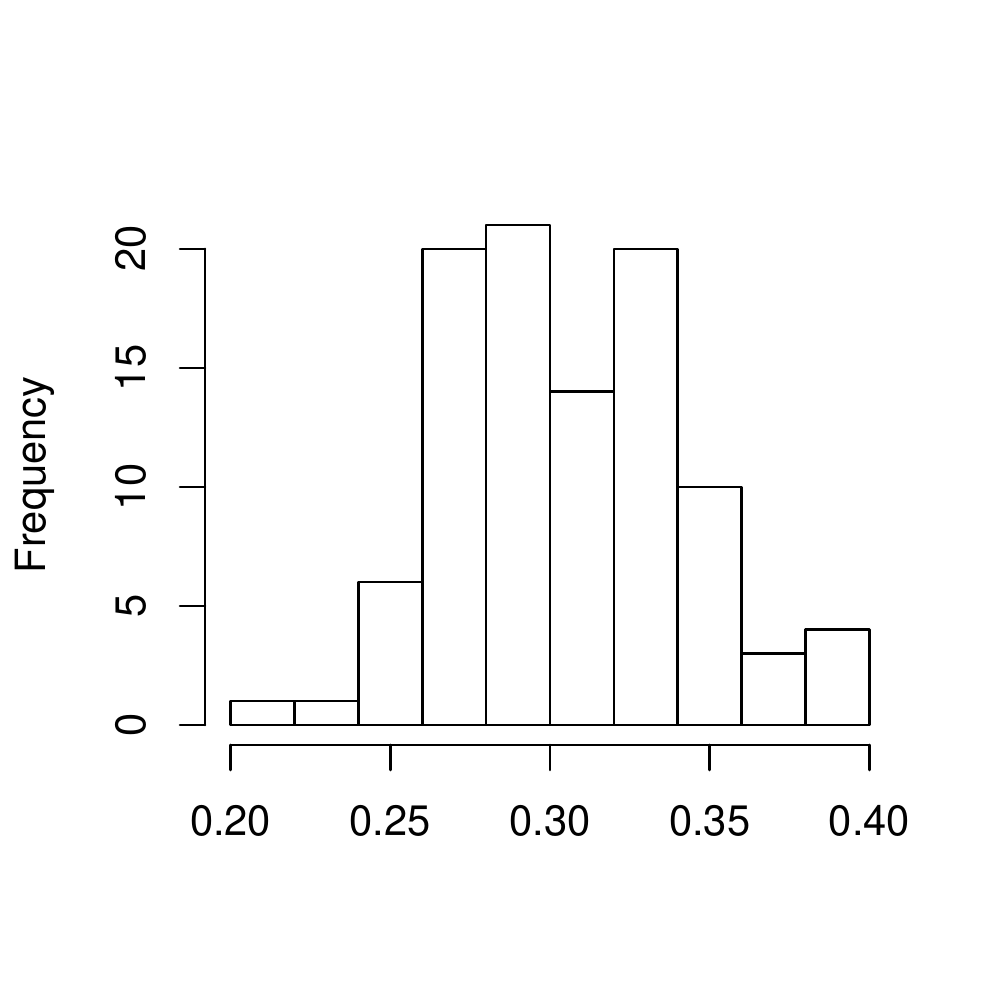}
\caption{Simulation Study II: Histogram of the ratios $d_i$/$e_i$ for all individuals from a randomly selected dataset under $N = 100$.}\label{fig:probit_hist}
\end{figure}

Finally, we examine the recovery of the individual latent curves, measured by the $L_2$ distance ratio $d_i/e_i$ defined in Study I. The EAP estimates of the individual curves are obtained by Monte Carlo approximation \eqref{eq:apprpost}, where $L = 100$ Monte Carlo samples are used. In particular, the histogram of $d_i/e_i$, $i = 1, ..., N$, is presented in Figure~\ref{fig:probit_hist}, for a randomly selected dataset among all replications under $N = 100$.  According to the histogram, $d_i$ is much smaller than $e_i$, though the ratios tend to be larger than those in Study I. It implies that, under the current setting,  the EAP estimate $\hat \theta_i(t)$ is still substantially more accurate than the population mean $\hat \alpha$ in estimating all individuals' latent curves.

\begin{table}
\begin{center}
\begin{tabular}{l cccccccccccccccccc}
\hline
 & $a_1$ & $a_2$ & $a_3$ & $a_4$ & $a_5$ \\
\hline
True & 1.00 & 1.00 & 0.65 & 0.62 & 0.53\\
\hline
MSE(N=50) & $\cdot$& $1.7\times 10^{-3}$& $7.9\times 10^{-4}$& $5.6\times 10^{-4}$& $4.1\times 10^{-4}$\\
MSE(N=100) &$\cdot$ & $7.5\times 10^{-4}$ & $3.2\times 10^{-4}$ & $2.7\times 10^{-4}$ & $2.1\times 10^{-4}$ \\
\hline
\hline
&$d_{1,1}$ & $d_{2,1}$ & $d_{3,1}$ & $d_{4,1}$ & $d_{5,1}$ \\
\hline
True& 0.00 & 0.45 & -0.25 & -0.27 & 0.34\\
\hline
MSE(N=50) & $\cdot$& $1.4\times 10^{-3}$& $6.2\times 10^{-4}$& $7.6\times 10^{-4}$& $6.5\times 10^{-4}$\\
MSE(N=100)& $\cdot$& $6.9\times 10^{-4}$& $3.3\times 10^{-4}$& $3.5\times 10^{-4}$& $3.8\times 10^{-4}$\\
\hline
\hline
&$d_{1,2}$ & $d_{2,2}$ & $d_{3,2}$ & $d_{4,2}$ & $d_{5,2}$ \\
\hline
True & 1.84 & 1.45 & 0.44 & 1.37 & 1.55\\
\hline
MSE(N=50) & $ 1.7 \times 10^{-3}$ & $2.1\times 10^{-3}$& $6.9\times 10^{-4}$& $1.0\times 10^{-3}$& $1.1\times 10^{-3}$ \\
MSE(N=100)& $6.7\times 10^{-4}$& $1.2\times 10^{-3}$& $4.3\times 10^{-4}$& $5.5\times 10^{-4}$& $5.4\times 10^{-4}$\\
\hline
\hline
& $\alpha$ & $\kappa$ & $c^2$ \\
\hline
True &  -0.79 & 0.30 & 1.27 \\
\hline
MSE(N=50) & $1.6\times 10^{-3}$ & $2.9\times 10^{-4}$ & $2.0\times 10^{-3}$ \\
MSE(N=100)  & $1.1\times 10^{-3}$ & $3.0\times 10^{-4}$ & $9.3\times 10^{-4}$ \\
\hline
\end{tabular}
 \caption{Simulation Study II: Simulation results on the parameter recovery accuracy for an LGP model with a probit measurement model component.}
\label{tab:sim3}
\end{center}
\end{table}

\section{Analysis of Negative Mood in BPD and MDD/DYS Patients}\label{sec:real}

We analyze data from a study of the affective instability in borderline personality disorder \citep{trull2008affective} that collected ecological momentary assessment data from psychiatric outpatients with borderline personality disorder (BPD) and with major depressive disorder (MDD) or dysthymic disorder (DYS).
The participants were recruited from one of four community mental health outpatient clinics through flyers.
The dataset has been analyzed in \cite{jahng2008analysis} and is downloaded from \url{http://dx.doi.org/10.1037/a0014173.supp}. The data contain  84 participants: 46 who met DSM-IV-TR \citep{edition2000diagnostic}
diagnostic criteria for BPD and who endorsed the diagnostic
feature of affective instability; and 38 who met DSM-IV-TR
diagnostic criteria for current MDD or DYS and did not report
affective instability.

\begin{figure}
  \centering
  \includegraphics[scale=0.8]{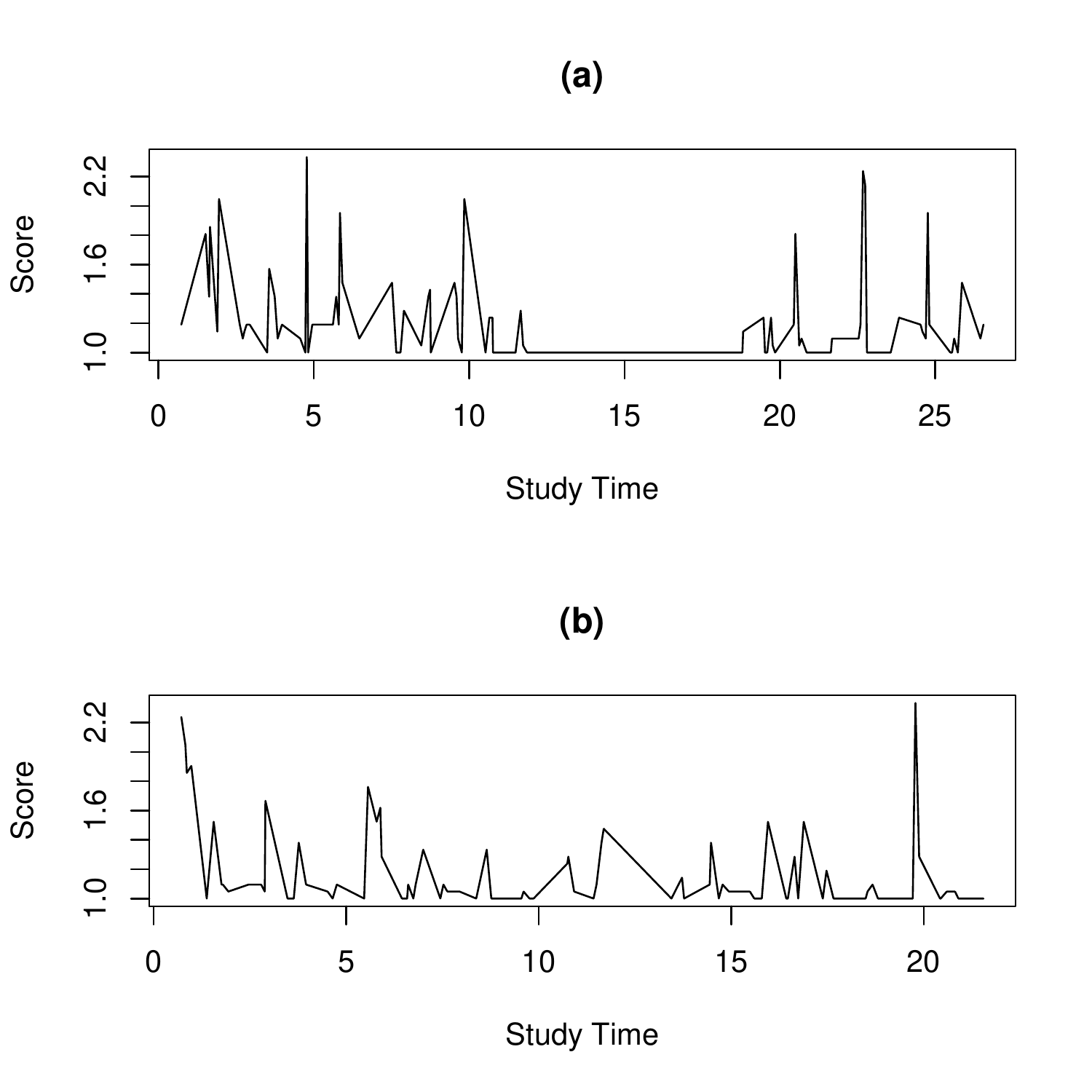}
  \caption{An illustration of the EMA data, where panels (a) and (b)
  show the negative affect composite score (y-axis) versus the study time (x-axis) from a MDD/DYS patient and a BPD patient, respectively.}\label{fig:data}
\end{figure}

This dataset contains, for each time and each participant, a negative affect composite score based on 21 items from the Positive and Negative Affect Scales-Extended Version \citep{watson1999panas}. The participants were measured multiple times a day over approximately 4 weeks of consecutive days.
As commonly encountered in EMA data,
the number of days of assessments per person and the
number of assessments per day differed (days per person:
median = 29, interquartile range = 2; assessments per
day: median =  5, interquartile range = 1). In total, the participants received
76 to 186 assessments (median =  153, interquartile range =  24) per person were conducted.
Table~\ref{tab:datastr} illustrates the data structure, where the five columns show the individual ID, the negative affect composite score, the group membership ($x_i = 0$ for the MDD/DYS group, $x_i = 1$ for the BPD group), the study time, and the calendar time, respectively. In particular, the study time uses day as the time unit and sets 00:00 of the first day receiving measurement as time 0 for each individual. Figure~\ref{fig:data} visualizes the data from a MDD/DYS patient and that from a BPD patient, where the individuals receive different numbers of measurement, at different and unequally spaced time points.

\begin{table}
  \centering
  \begin{tabular}{ccccc}
    \hline
    ID& Score& Group& Study Time& Calendar Time\\
    \hline
1&1.19& 0&0.74 &2005-03-18 17:40:00 \\
1&1.81& 0&1.52  &2005-03-19 12:24:38 \\
1&1.38& 0&1.63  &2005-03-19 15:06:36 \\
1&1.86& 0&1.66  &2005-03-19 15:49:34 \\
$\vdots$&$\vdots$&$\vdots$&$\vdots$&$\vdots$\\
\hline
  \end{tabular}
  \caption{An illustration of the EMA data from the mood study of BPD and MDD/DYS patients.}\label{tab:datastr}
\end{table}

Following the research question of \cite{jahng2008analysis}, we investigate, by making use of the proposed latent Gaussian process model, whether the BPD group suffers from 
more temporal negative mood instability than the MDD/DYS group.
We also investigate the mean of the negative mood of the two groups.
To answer these questions under the latent Gaussian process modeling framework,
we treat the negative affect composite score as a continuous variable and adopt a single-indicator linear factor measurement model. In addition, we assume the mean and the kernel functions of the latent Gaussian process are group specific. 
Specifically, the model is specified as follows.
\begin{equation*}
\begin{aligned}
Y_{i1}(t)\vert \theta_i(t) &\sim N(\theta_{i}(t), \sigma^2),\\
\theta_i(\cdot)\vert x_i &\sim GP(m_{x_i}, K_{x_i}),
\end{aligned}
\end{equation*}
where $m_0(t) = \alpha_0$, $m_1(t) = \alpha_1$, $K_0(t,t') = c_0^2 \exp\left({-(t-t')^2}/{(2\kappa_0^2)}\right)$, and
$K_1(t,t') = c_1^2 \exp\left({-(t-t')^2}/{(2\kappa_1^2)}\right)$. Under these assumptions, the Gaussian process for each group is stationary. According to the recruitment design of the study, the stationarity assumption seems reasonable.

The main results are shown in Table~\ref{tab:realdata1}, including parameter estimates obtained from the EM algorithm and their $95\%$ bootstrap confidence interval \citep[Chapter 6,][]{efron1994introduction}. The bootstrap results are obtained by resampling individuals with replacement. In particular, an estimate of the variance due to the measurement error is $\hat \sigma^2 = 0.091$, which is much smaller than $\hat c_0^2 = 0.234$ and $\hat c_1^2 = 0.440$, the overall variations of the two Gaussian processes. In addition, the two groups only significantly differ by the overall long-run variations, with a difference $\hat c_1^2 - \hat c_0^2 = 0.206$ which has a corresponding 95\% bootstrap confidence interval $(0.010, 0.407)$. That is, the BPD group has more variation in the long run than the MDD/DYS group, which is consistent with the existing knowledge these mental health disorders.
Their overall mean scores are not significantly different, for which the difference is $\hat \alpha_1 - \hat \alpha_0 = 0.081$ and a 95\%  confidence interval $(-0.106, 0.275)$. Similarly, the two groups do not significantly differ in terms of the short-term temporal dependence, evidenced by $\hat \kappa_1 - \hat \kappa_0 = 0.012$ and its 95\%  confidence interval $(-0.039, 0.060)$.

\begin{figure}
\centering
\begin{subfigure}{.5\textwidth}
  \centering
  \includegraphics[width=1\linewidth]{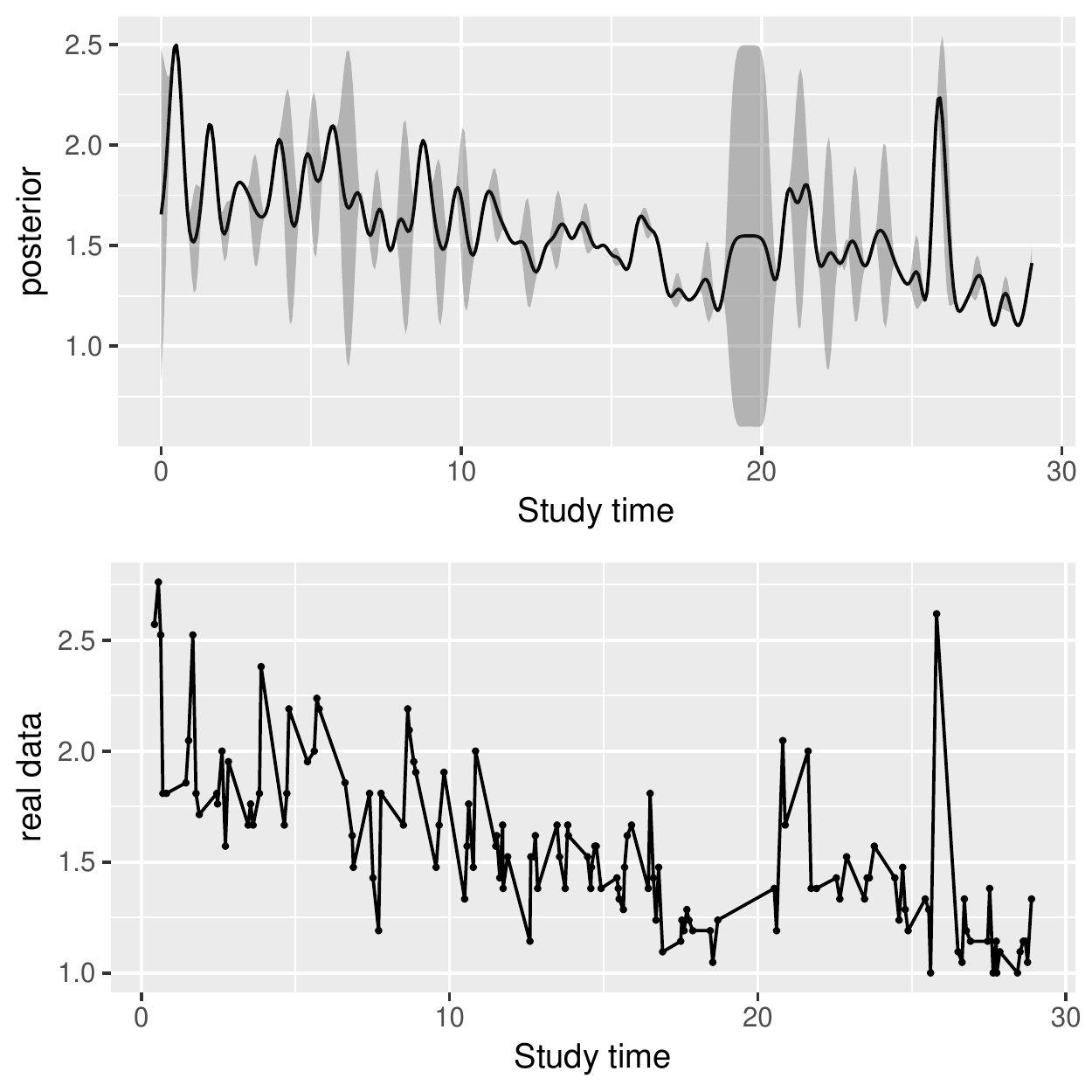}
  \caption{Participant 47 (from the MDD/DYS group)}
\end{subfigure}%
\begin{subfigure}{.5\textwidth}
  \centering
  \includegraphics[width=1\linewidth]{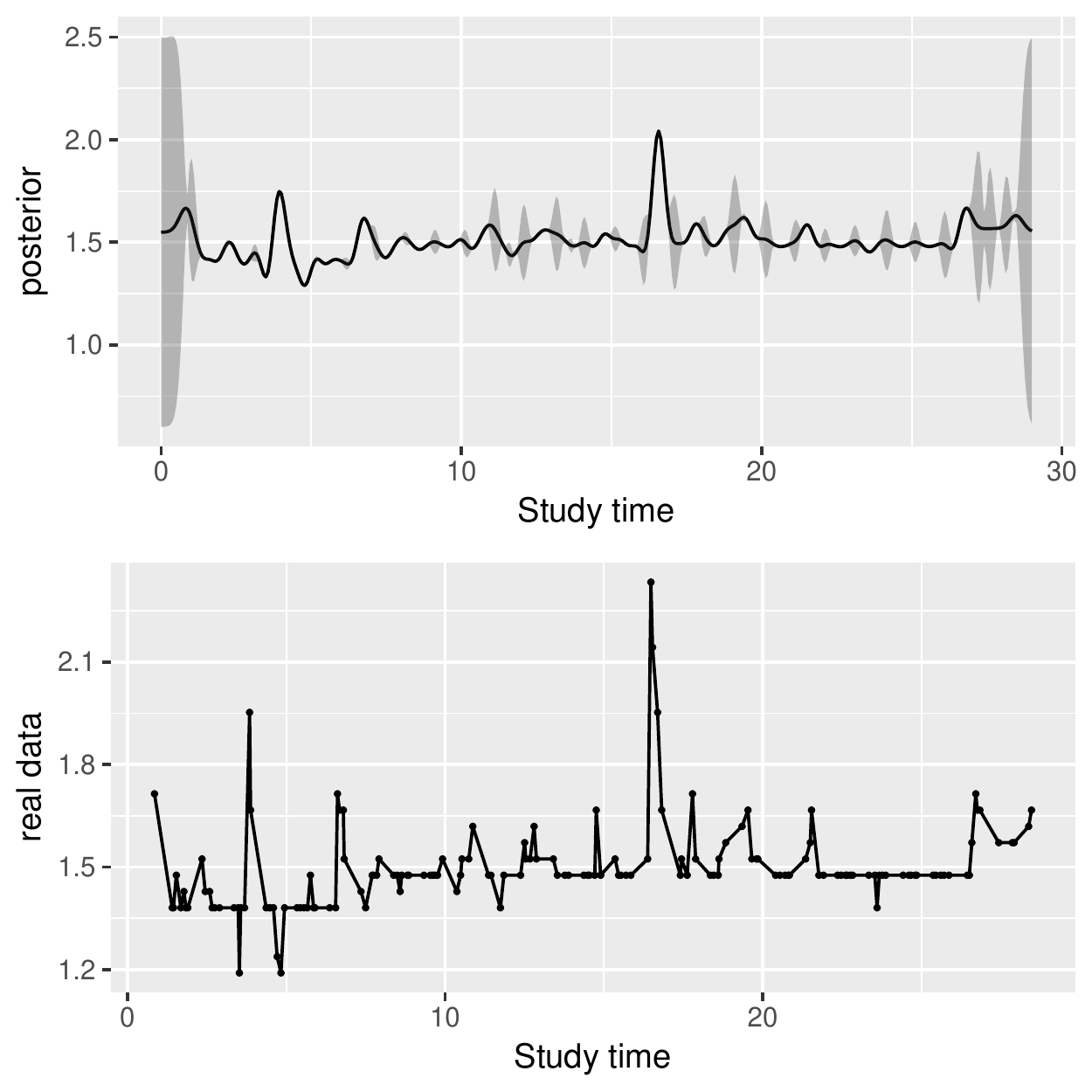}
  \caption{Participant 17 (from the MDD/DYS group)}
\end{subfigure}
\\
\centering
\begin{subfigure}{.5\textwidth}
  \centering
  \includegraphics[width=1\linewidth]{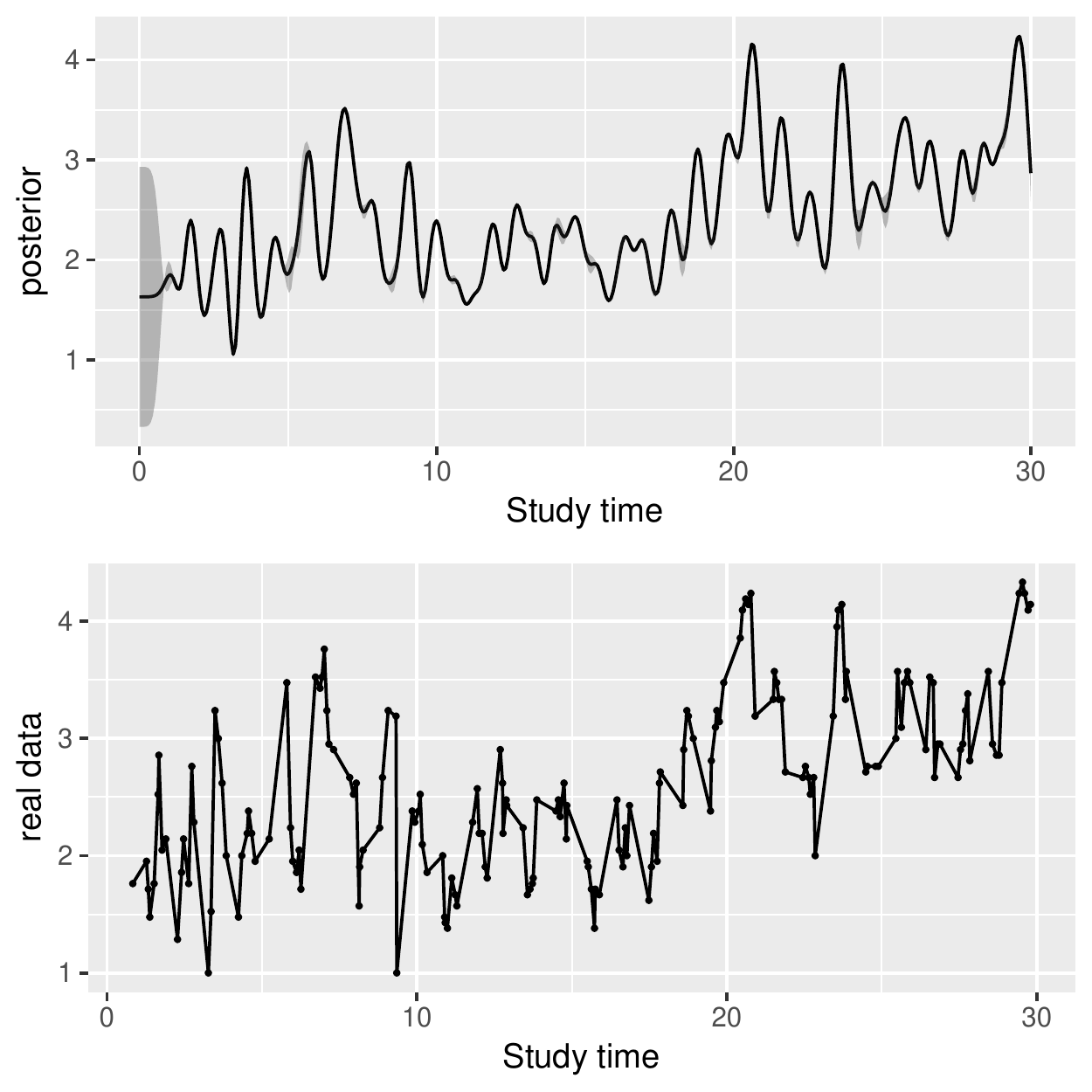}
  \caption{Participant 9 (from the BPD group)}
\end{subfigure}%
\begin{subfigure}{.5\textwidth}
  \centering
  \includegraphics[width=1\linewidth]{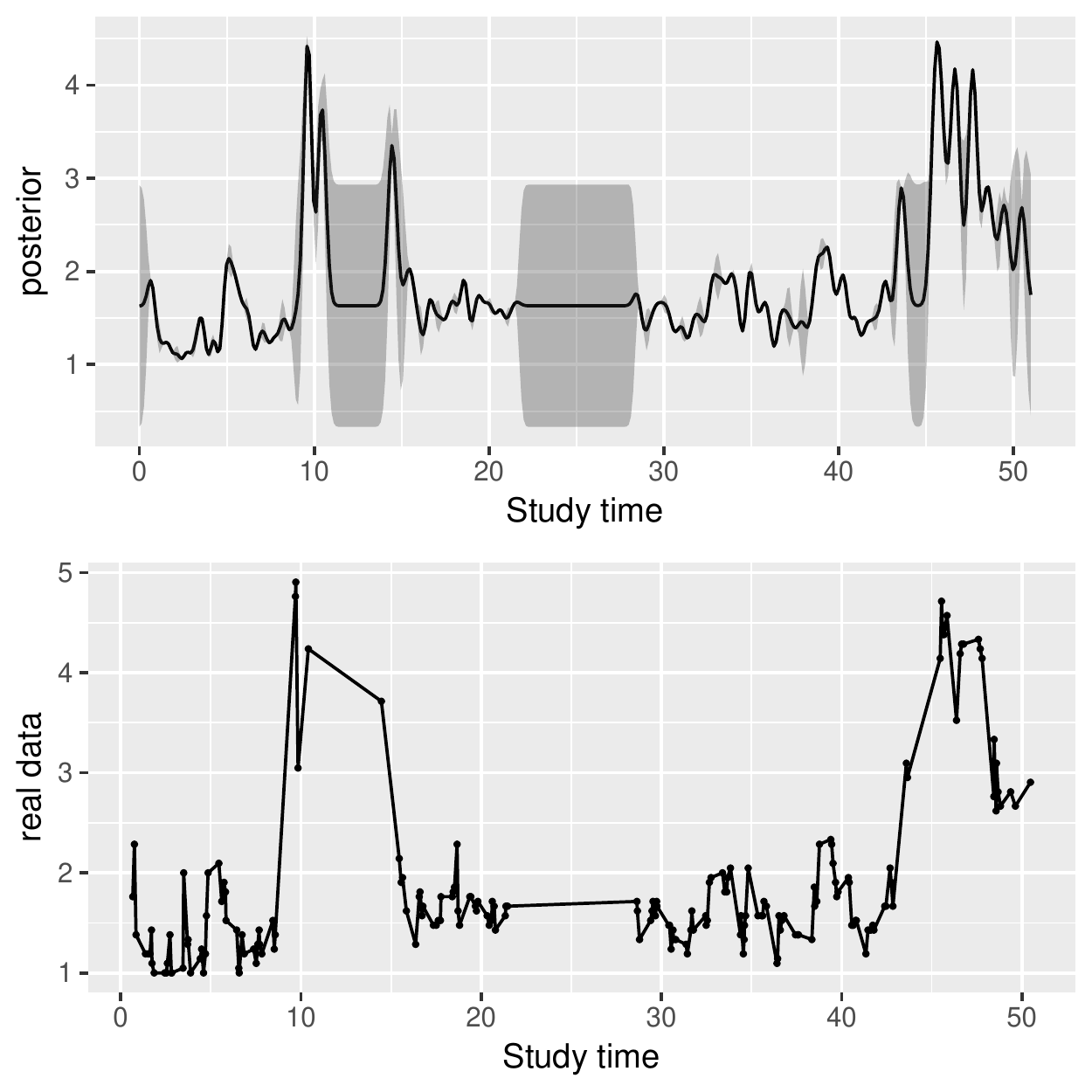}
  \caption{Participant 35 (from the BPD group)}
\end{subfigure}
    \caption{The posterior mean and the posterior $2.5\%$ and $97.5\%$ quantiles of $\theta_i(t)$, as well as the corresponding response process, of four participants. Participants in panels (a) and (b) are from the MDD/DYS group and participants in panels (c) and (d) are from the BPD group.}\label{fig:1}
\end{figure}

In addition to the estimation of the model parameters, the proposed modeling framework allows us to make inference at the individual level. To demonstrate, in Figure~\ref{fig:1}, we show the posterior mean and the posterior $2.5\%$ and $97.5\%$ quantiles of $\theta_i(t)$, as well as the corresponding response process, of four participants, two of whom are from the MDD/DYS group and the other two from the BPD group. The calculation of the posterior mean and the posterior quantile for $\theta_i(t)$ is described in Section~\ref{sec:estimation}. As we can see, the posterior mean of $\theta_i(t)$ is quite smooth and captures the overall trend of the response process. In addition, the confidence band, given by the posterior $2.5\%$ and $97.5\%$ quantiles of $\theta_i(t)$, becomes wide when two subsequent measurements have a long time lag. For example, participant 35 from the BPD group did not have measurement from
the 11th to the 13th day and from the 22nd to the 27th day. That is why the wide confidence bands are observed in panel (d) within the corresponding intervals.
When there are multiple measurements occur around a single time point $t$, the posterior variance at time $t$ can be close to 0 and consequently the corresponding  posterior mean and posterior $2.5\%$ and $97.5\%$ quantiles are close to each other.

\begin{table}
  \centering
  \begin{tabular}{ccccccccc}
    \hline
     & $\hat \alpha_0$& $\hat \alpha_1$ & $\hat c_0^2$ & $\hat c_1^2$ &$\hat \kappa_0$ & $\hat \kappa_1$ & $\hat \sigma^2$ \\
    \hline
      Point estimate  & 1.549 & 1.630 & 0.234 & 0.440 & 0.237 & 0.249 & 0.091 \\

      95\% CI lower bound  & 1.436 & 1.476 & 0.154 & 0.273 & 0.192 & 0.201 & 0.071\\
      95\% CI upper bound  & 1.658 & 1.793 & 0.304 & 0.608 & 0.272 & 0.281 & 0.112\\
    \hline
  \end{tabular}
  \caption{Results from fitting an LGP model to the EMA data  from a mood study of BPD and MDD/DYS patients.}\label{tab:realdata1}
\end{table}

\section{Concluding Remarks}\label{sec:conclu}

In this paper, we introduce the latent Gaussian process model as a general family of continuous-time latent curve models. This new model complements the existing models for the analysis of intensive longitudinal data.
The proposed model decomposes the latent curve analysis into a measurement model component and a structural model component. The measurement component captures the conditional distribution of an individual's observed data given his/her latent curve in a continuous time domain and 
the structural component models the distribution of the latent curve. It is shown that many existing latent curve models are special cases of the proposed one.

In particular, a Gaussian process model is proposed for the modeling of latent curves in the structural model component. By making use of the mathematical properties of Gaussian processes, the modeling of the structural component is further decomposed into separate modeling of the mean function and the Kernel function of a Gaussian process.
Estimation and statistical inference are further discussed under an empirical Bayes framework, where inference is considered at both population and individual levels.

The proposed model and methods are further illustrated through simulation studies and a real data example. In particular, our analysis of the negative mood of BPD and MDD/DYS patients reveals that the main difference between the two groups is due to the BPD group having significantly higher
long-term variation, while the two groups are not significantly different in the mean negative affect levels and in the short-term temporal dependence.

The proposed framework leads to many new directions, which are left for future investigation.  First, it is often of interest to measure multiple correlated dynamic latent traits, in which case $\theta_i(t)$ becomes a vector at each time point $t$. The current framework can be easily extended to that setting, by adopting a multidimensional measurement model (e.g., multidimensional item response theory model) and a multivariate Gaussian process model for the structural component.
Second, many intensive longitudinal studies involve not only measurement but also interventions (e.g., treatment of mental health disorders). Interventions can be viewed as time-dependent covariates which can be incorporated into the structural component of the proposed model. By estimating the coefficients associated with the intervention covariates,
the intervention effects can be evaluated dynamically.
Finally, the psychometric properties of the proposed model remain to be studied,
such as the detection of differential item functioning, the assessment of model goodness-of-fit, and the evaluation of measurement reliability.

\bibliographystyle{apacite}

\end{document}